\documentclass[conference]{IEEEtran}

\newif\iflongver\longvertrue 

\iflongver
\IEEEoverridecommandlockouts 
\else
\fi

\usepackage{cite}

\usepackage{latexsym,amssymb}

\usepackage{hyperref}
\usepackage{latexsym,amssymb} 
\usepackage{color} 
\usepackage{listings} 
\usepackage{mathpartir} 
\usepackage[only,Mapsto,rightarrowtriangle,llbracket,rrbracket]{stmaryrd} 
\usepackage{relsize}
\usepackage{tikz}
\usetikzlibrary{automata,positioning,arrows}

\usepackage{amsthm} 

\newcommand{\dt}[1]{\textbf{\emph{#1}}} 

\newcommand{\sep}{\mathbin{\mid}} 
\newcommand{\Sep}{\mathbin{\:\mid\:}} 

\newcommand{\prefixeq}{\preceq} 
\newcommand{\prefix}{\prec} 

\newcommand{\traceSim}{\asymp} 

\renewcommand{\neg}{\mbox{\relsize{-1}$\lnot$}}

\newcommand{\union}{\cup}
\newcommand{\intersect}{\cap}
\newcommand{\proves}{\vdash}
\newcommand{\aftqua}{.\:}
\newcommand{\all}[2]{\forall #1 \aftqua #2}
\newcommand{\some}[2]{\exists #1 \aftqua #2}

\newcommand{\nat}{\mathbb{N}} 

\newcommand{\imp}{\Rightarrow}

\newcommand{\segs}{\mathconst{segs}}
\newcommand{\ok}{\mathconst{ok}}
\newcommand{\fst}{\mathconst{fst}}
\newcommand{\snd}{\mathconst{snd}}
\newcommand{\lab}{\mathconst{lab}}
\newcommand{\labs}{\mathconst{labs}}
\newcommand{\sub}{\mathconst{sub}}
\newcommand{\cpath}{\mathconst{cpath}}
\newcommand{\fsuc}{\mathconst{fsuc}}
\newcommand{\elab}{\mathconst{elab}}
\newcommand{\sameCtl}{\mathconst{sameCtl}}
\newcommand{\sameExcept}{\mathconst{sameExcept}}
\newcommand{\cfge}{\mathord{\rightarrow}} 

\newcommand{\eplus}{{\mbox{\relsize{-4}$\oplus$}}}

\newcommand{\ctrl}{\mathconst{ctrl}} 
\newcommand{\store}{\mathconst{stor}}

\newcommand{\Left}{\mathconst{left}} 
\newcommand{\Right}{\mathconst{right}}

\newcommand{\LT}{\mathconst{lt}} 
\newcommand{\RT}{\mathconst{rt}}

\newcommand{\destutter}{\mathconst{destutter}}
\newcommand{\CFG}{\mathconst{CFG}}
\newcommand{\PRE}{\mathconst{pre}}   
\newcommand{\POST}{\mathconst{post}} 

\newcommand{\Var}{\mathconst{Var}} 

\newcommand{\aut}{\mathconst{aut}}

\newcommand{\last}[1]{#1_{-1}} 
\newcommand{\init}{\mathit{init}}
\newcommand{\fin}{\mathit{fin}}
\newcommand{\trans}{\mapsto} 
\newcommand{\tranSeg}[1]{\stackrel{#1}{\longmapsto}} 
\newcommand{\biTrans}{\Mapsto} 

\newcommand{\ctrans}{\rightarrowtriangle} 
\newcommand{\config}[2]{\langle #1,\: #2\rangle}

\newcommand{\uequiv}{\mathrel{\cong}} 

\newcommand{\mathconst}[1]{\mbox{\relsize{-1}\textsf{#1}}} 
\newcommand{\rn}[1]{\textsc{\relsize{-1}#1}} 

\newcommand{\update}[3]{[#1\, |\, #2\scol\, #3]} 
\newcommand{\scol}{\mathord{:}} 

\newcommand{\leftex}[1]{ \raisebox{.25ex}{\relsize{-1}$\langle\hspace*{-2.1pt}[$} #1 \raisebox{.25ex}{\relsize{-1}$\langle\hspace*{-2.0pt}]$} }
\newcommand{\rightex}[1]{ \raisebox{.25ex}{\relsize{-1}$[\hspace*{-2.0pt}\rangle$} #1 \raisebox{.25ex}{\relsize{-1}$]\hspace*{-2.1pt}\rangle$} }
\newcommand{\leftF}[1]{\leftex{#1}}
\newcommand{\rightF}[1]{\rightex{#1}}

\newcommand{\eqbi}[2]{#1\mathrel{\ddot{=}}#2} 
\newcommand{\eqbib}[2]{#1\mathrel{\mathring{=}}#2} 

\newcommand{\keyw}[1]{\ensuremath{\mathsf{#1}}} 

\newcommand{\skipc}{\keyw{skip}}
\newcommand{\ifc}[3]{\keyw{if}\ {#1}\ \keyw{then}\ {#2}\ \keyw{else}\ {#3}\ \keyw{fi}}
\newcommand{\whilec}[2]{\keyw{while}\ {#1}\ \keyw{do}\ {#2}\ \keyw{od}}

\newcommand{\lskipc}[1]{\keyw{skip}^{#1}}
\newcommand{\lassg}[3]{#2:=^{#1}#3}
\newcommand{\lifc}[4]{\keyw{if}^{#1}\ {#2}\ \keyw{then}\ {#3}\ \keyw{else}\ {#4}\ \keyw{fi}}
\newcommand{\lwhilec}[3]{\keyw{while}^{#1}\ {#2}\ \keyw{do}\ {#3}\ \keyw{od}}

\newcommand{\lchoice}[3]{#2\mathrel{\sqcup}^{#1}#3} 
\newcommand{\choice}[2]{#1\mathrel{\sqcup}#2} %

\renewcommand{\P}{\mathcal{P}}  
\renewcommand{\S}{\mathcal{S}} 
\newcommand{\Lrel}{\mathcal{L}} 
\newcommand{\Q}{\mathcal{Q}} 
 
\newcommand{\R}{\mathcal{R}} 

\newcommand{\Z}{\mathbb{Z}} 

\newcommand{\mod}{\mathbin{\%}}

\newcommand{\Dot}[1]{{#1}^{\bullet}}
\newcommand{\fdot}{\mathconst{dot}} 
\newcommand{\fdotsmall}{\mbox{\relsize{-1}\mathconst{dot}}} 
\newcommand{\subst}[3]{{#1}^{#2}_{#3}}

\definecolor{light-gray}{gray}{0.88}
\definecolor{dark-gray}{gray}{0.25}
\newcommand{\ghostbox}[1]{\colorbox{light-gray}{#1}} 

\newcommand{\specSym}{\leadsto}
\newcommand{\rspecSym}{\ensuremath{\mathrel{\mbox{\footnotesize$\thickapprox\hspace*{-.4ex}>$}}}}

\newcommand{\spec}[2]{#1\specSym #2} 
\newcommand{\rspec}[2]{#1\rspecSym #2} 

\definecolor{dkgreen}{rgb}{0,0.6,0}

\lstdefinelanguage{TOY}
{
basicstyle=\normalsize\rmfamily\itshape, 
keywordstyle=[1]\upshape\color{blue}\bf,%
morekeywords=[1]{while,do,od,%
var,if,then,else,fi,skip%
},%
string=[b]",%
commentstyle=\itshape\color{dkgreen},%
columns=[l]fullflexible,%
sensitive=true,%
morecomment=[s]{(*}{*)},%
numberstyle=\upshape,
keepspaces=true,%
literate=%
 {:=}{{$:=$}}1  
 {\\expx}{{$2^x$}}1  
 {\\uequiv}{{$\:\uequiv\:$}}1  
 {*<}{$\lexop$}{1}
 {*]}{$\lexcl$}{1}%
 {*>}{$\rexcl$}{1}%
 {*[}{$\rexop$}{1}%
 {=:=}{$\eqbi$}{1}%
 {<}{$<$}{1}%
 {>}{$>$}{1}%
 {-}{$\,-$}{1}%
 {'}{$\,^\prime$}{1}%
 {<=}{$\le$}{1}%
 {>=}{$\ge$}{1}%
 {<>}{$\ne$}{1}%
 {/\\}{$\land$}{1}%
 {mod}{$\mod$}{1}%
 {\\/}{ $\lor$ }{3}%
 {not\ }{$\lnot$ }{2}%
}

\newenvironment{ditemize}{
\begin{list}{$\bullet$}{%
\setlength{\itemsep}{0pt}\setlength{\rightmargin}{0pt}%
\setlength{\leftmargin}{.6em}\setlength{\parsep}{0ex}}}{
\end{list}}

\newenvironment{dlist}{%
\begin{list}{}{%
\setlength{\itemsep}{0pt}\setlength{\rightmargin}{0pt}%
\setlength{\leftmargin}{1em}\setlength{\parsep}{0ex}}}{
\end{list}}

\newtheorem{theorem}{Theorem}
\newtheorem{lemma}[theorem]{Lemma}
\newtheorem{proposition}[theorem]{Proposition}

\begin{document}

\title{Alignment Completeness for \\
Relational Hoare Logics
\iflongver (with appendix)\fi
}

\author{\IEEEauthorblockN{Ramana Nagasamudram}
\IEEEauthorblockA{Stevens Institute of Technology}
\and
\IEEEauthorblockN{David A. Naumann}
\IEEEauthorblockA{Stevens Institute of Technology}
}

\IEEEoverridecommandlockouts
\IEEEpubid{\makebox[\columnwidth]{978-1-6654-4895-6/21/\$31.00~
\copyright2021 IEEE \hfill} \hspace{\columnsep}\makebox[\columnwidth]{ }}
\maketitle

\begin{abstract} 
Relational Hoare logics (RHL) provide rules for reasoning about relations between programs.  Several RHLs include a rule we call sequential product that infers a relational correctness judgment from judgments of ordinary Hoare logic (HL). Other rules embody sensible patterns of reasoning and have been found useful in practice, but sequential product is relatively complete on its own (with HL). As a more satisfactory way to evaluate RHLs, a notion of alignment completeness is introduced, in terms of the inductive assertion method and product automata. Alignment completeness results are given to account for several different sets of rules. The notion may serve to guide the design of RHLs and relational verifiers for richer programming languages and alignment patterns.
\end{abstract}


\section{Introduction}\label{sec:intro}

A common task in programming is to ascertain whether a modified version of a program is equivalent to the original.  For programs with deterministic results, equivalence can be formulated simply: From any initial state, if both programs terminate then their final states are the same.
This termination-insensitive property is akin to partial correctness:
A program $c$ satisfies $\spec{P}{Q}$ if its terminating executions, from initial states 
that satisfy $P$, end in states that satisfy $Q$.  To relate programs $c$ and $d$, use binary relations $\R,\S$ over states: $c$ and $d$ satisfy $\rspec{\R}{\S}$ if pairs of their terminating executions, from $\R$-related initial states, end in $\S$-related states. 
For program equivalence, take $\R$ and $\S$ to be the identity on states.  

Hoare logics (HL) provide sound rules to infer correctness judgments $c:\spec{P}{Q}$, conventionally written $\{P\}\,c\,\{Q\}$, for various programming languages.  In this paper we confine attention to simple imperative commands and focus on Relational Hoare logics (RHL) which provide rules to
infer judgments which we write as $c\sep d: \rspec{\R}{\S}$.  Such properties go beyond
program equivalence. Using primed variables to refer to the second of two related states, the judgment
$c\sep d:\rspec{x=x'}{y<y'}$ expresses that when run from states that agree on the initial value of $x$,
the final value of $y$ produced by $c$ is less than the final value of $y$ from $d$
(i.e., $d$ majorizes $c$).
As another example, $c\sep c:\rspec{x=x'}{y=y'}$ relates $c$ to itself and says that the final value of $y$ is determined by the initial value of $x$.  Such dependency properties arise in many contexts including compiler optimization and security analysis which motived early work on RHL~\cite{Benton:popl04}.

Suppose the variables of $d$ are disjoint from those of $c$. For example let $d$ be a copy of $c$
with all the variables renamed by adding primes.  Then a relation on states amounts to a predicate on primed and unprimed variables as in the preceding informal notation.  
Moreover terminated executions of $(c;d)$ are in bijection with pairs of terminated executions of $c$ and $d$. Put differently, $c;d$ serves as a product program, much like products in automata theory.
The product program lets us prove relational properties using HL, but in general the 
technique is unsatisfactory.  As an example, consider the simple program $c0$ in Fig.~\ref{fig:examples} which computes in $z$ the factorial of $x$.  Let $c0'$ be a renamed copy,
so determinacy of $c0$ can be expressed by $\spec{x=x'}{z=z'}$.  
To prove the judgment 
$c0;c0': \spec{x=x'}{z=z'}$
in HL 
we need the assertion $z=x!\land x=x'$ at the semicolon,
to get $z=x!\land z'=x'!\land x=x'$ following $c0'$, from which $z=z'$ follows.
The spec $\spec{x=x'}{z=z'}$ involves nothing more than equalities, yet the proof
requires nonlinear arithmetic. This illustrates the general problem that the technique
requires strong functional properties and thus nontrivial loop invariants
(as famously observed in~\cite{TerauchiAiken}).

\begin{figure*}[t]    
\begin{lstlisting}[basicstyle=\footnotesize\rmfamily\itshape] 
                       c0: (* z := x! *) y:= x; z:= 1; while y <> 0 do z:= z*y; y:= y-1 od
                       c1: (* z := \expx *) y:= x; z:= 1; while y <> 0 do z:= z*2; y:= y-1 od
                       c2: (* z := x! *) y:= x; z:= 1; w:= 0; while y <> 0 do if w mod 2 = 0 then z:= z*y; y:= y-1 fi;  w:= w+1 od
                       c3: (* z := \expx *) y:= x; z:= 1; w:= 0; while y <> 0 do if w mod 3 = 0 then z:= z*2; y:= y-1 fi; w:= w+1 od
\end{lstlisting}
\vspace*{-1ex}
\caption{Example programs (where $\mod$ is modulo). 
}\label{fig:examples}
\end{figure*}

There is a simple way to prove $c0\sep c0': \rspec{x=x'}{z=z'}$.
Consider the execution pairs to be aligned step-by-step, and note that at each aligned pair of states we have $y=y'\land z=z'$, given $x=x'$ initially.  
RHLs feature rules for compositional reasoning about similarly-structured 
subprograms, 
which embody informal patterns of reasoning and enable the use of simple 
assertions with alignment of computations expressed in terms of syntax.
For example, from judgments $y:=x\sep y':=x' : \rspec{x=x'}{y=y'}$ 
and $z:=1\sep z':=1 :\rspec{y=y'}{(y=y'\land z=z')}$ infer 
that 
$y:=x;z:=1 \sep y':=x'; z':=1 :\rspec{(x=x')}{(y=y'\land z=z')}$.
A number of RHLs have appeared in the literature, but even for the simple imperative 
language we see no convergence on a common core set of rules. Besides 
closely ``synchronized'' rules like the sequence rule for the preceding inference (see \rn{dSeq} in Fig.~\ref{fig:lockstep}), there are sound rules to relate a command to skip  (Fig.~\ref{fig:one-side})
or to a differently-structured command.
For an example of the latter see Sec.~\ref{sec:discuss}.

The touchstone for Hoare logics is Cook's completeness theorem~\cite{Cook78} which says 
any true correctness judgment $c:\spec{P}{Q}$ can be derived using the rules. 
To be precise, HL relies on entailments between assertions
and the rules are complete \emph{relative} to completeness of assertion reasoning.
Moreover, completeness requires \emph{expressiveness} of the assertion language, meaning
that weakest preconditions can be expressed, for whatever types of data are manipulated
by the program~\cite{AptOld3}.  
These considerations are not important for the
ideas in this paper.  We follow the common practice of treating assertions and their entailments 
semantically~\cite{NipkowCSL02} (i.e., by shallow embedding in our metalanguage).

RHLs often feature a rule we dub ``sequential product'', which infers $c\sep c':\rspec{\R}{\S}$
from the HL judgment $c;c' :\spec{\R}{\S}$, formalizing the idea of product program.  
It is a useful rule, to complement the rules for relational judgments about similarly-structured programs. For example, in the regression verification~\cite{GodlinS09} scenario with which we began,
i.e., equivalence between two versions of a program,
same-structure rules can be used to relate the unchanged parts, while sequential product can be used for the revised parts if they differ in control structure.
This is how programmers might reason informally about a small change in a big program.

But there is a problem.  Consider the logic comprised of the sequential product rule together with a sound and complete HL.   This is complete, in the sense of Cook, for relational judgments!  
The problem is well known to RHL experts.  
Completeness is the usual means to
determine a sufficient and parsimonious set of inference rules, but completeness
fails to discriminate between a RHL that supports compositional proofs using facts about aligned subprograms and an impoverished RHL with only sequential product.

\emph{The \underline{conceptual contribution} of this paper is the notion of alignment completeness,
which discriminates between rules in terms of different classes of alignments.
The \underline{technical contributions} are four alignment completeness theorems, for
representative collections of RHL rules.}

To explain the idea our first step is to revisit Floyd-Hoare logic. 
Floyd~\cite{Floyd67} made precise the 
inductive assertion method (IAM) already evident in
work by Turing~\cite{TuringChecking49} (see~\cite{AptO19}).
To prove $\spec{P}{Q}$ by IAM, provide assertions at control points in the program,
at least $P$ at the initial point and $Q$ at program exit.
We call this an \dt{annotation}; it is familiar to programmers in the form of assert statements,
and it gives rise to verification conditions.
A \dt{valid} annotation is one where the verification conditions are true
and every loop in control flow is ``cut'' by an annotation. 
A valid annotation constitutes a proof by induction.    
HL is complete in a sense that we call \dt{Floyd completeness}:
\emph{For any valid annotation of a program $c$ for a spec $\spec{P}{Q}$,
there is a proof in HL of $c: \spec{P}{Q}$ using only judgments
of the form $b:\spec{R}{S}$ with $b$ ranging over subprograms of $c$
and $R,S$ the assertions annotating the entry and exit points of $b$.}\footnote{The precise result depends on details of the HL rules and may require mildly adjusted judgments in addition to those
directly given by the annotation; see Thm.~\ref{thm:floydComplete}.}

Now consider relating two programs.  An annotation should attach relations to designated pairs of points in the control flow of the two programs.
For the example judgment $c0\sep c0': \rspec{x=x'}{z=z'}$, choose pairs at the lockstep positions,
and the abovementioned conjunction $y=y'\land z=z'$.
Validity of an annotation is defined in terms of execution pairs aligned in accord with 
the designated pairs of control points: at aligned steps, the asserted relations hold.
To make alignment precise we use product automata.
A product represents a particular pattern of alignment; if it is adequate in the 
sense of covering all execution pairs then the IAM can be applied to prove 
relational properties of the two programs.
A set of RHL rules is then \dt{alignment complete}, for a given class of alignment automata,
if \emph{for any valid annotated automaton there is a derivation using
the rules and only the assertions and judgments associated with the annotated automaton.}

This paper formalizes the idea and gives some representative results of this kind:
for sequential product, for strict lockstep, and also for data-dependent alignment of loop iterations.
To see the need for the latter, consider program $c2$ in Fig.~\ref{fig:examples},
in which some iterations have no effect on $y$ or $z$.
We can prove 
$c0\sep c2: \rspec{x=x'}{z=z'}$
using only simple equalities and without reasoning about factorial, provided the effectful iterations are aligned in lockstep while the gratuitous iterations
of $c2$ (when \lstinline{w} is odd) proceed with $c0$ considered to be stationary.

The notion of Floyd completeness should be no surprise to readers familiar with Floyd-Hoare logic or related topics like software model checking.
But the authors are unaware of any published result of this form. 
A related idea is proof outline logic~\cite{OwickiGries,AptOld3},
which formalizes commands with correct embedded assertions.  Rules for proof outlines have  
verification conditions which imply an annotation is valid.  
In addition to showing Cook-style soundness and completeness results for a proof
outline logic, Apt et al.\ prove a ``strong soundness'' theorem~\cite[Thm.~3.3]{AptOld3} 
which says that if the program's proof outline is provable then in any execution each assertion is true when control is at that point.  
The converse would be a way to formalize Floyd completeness.
Strong soundness is phrased in terms of transition semantics (small steps).
By contrast, the fact that reasoning in HL is compositional in terms of control structure is
beautifully reflected in proofs of soundness and (Cook) completeness based
on denotational semantics~\cite{AptOld3}.

In this paper we only consider rules that are sound, in the usual sense,
and we have no need to formalize alignment soundness.

\subsubsection*{Outline}

Sec.~\ref{sec:prelim} lays the groundwork by spelling out Floyd completeness for HL,
in terms of automata with explicit control points, including automata based on small-step semantics of labelled commands.   
Sec.~\ref{sec:prodAut} formalizes product automata: ways in which a pair of automata can be combined into an automaton on pairs of states that serves to represent aligned steps of two computations.

Sec.~\ref{sec:seqcomplete} gives the first alignment completeness theorem:
Given a valid annotation of a product automaton for 
$c|c': \rspec{P}{Q}$
that executes $c$ to termination and then executes $c'$, 
there is a proof using just the sequential product rule and the rules of HL ---and using only judgments for subprograms with pre- and postconditions given by the annotation.

Sec.~\ref{sec:lockstep} gives the alignment completness theorem for lockstep alignment: if $c\sep c': \rspec{P}{Q}$ is witnessed by such an alignment with valid annotation,
then it can be proved without HL, using just the RHL rules that relate same-structured programs 
(see Fig.~\ref{fig:lockstep}).  And again the judgments used are those associated
with the annotation.
These rules are not complete in the usual sense, but lockstep reasoning 
is sufficient in some practical situations.

The theorem of Sec.~\ref{sec:lockstep-seq} accounts for the combination of \rn{SeqProd} 
with the lockstep RHL rules.  This and the preceding results are for alignments that can be 
described in terms of which control points are aligned.
Sec.~\ref{sec:dissonant} accounts for conditional alignment of loop iterations,
using a rule due to Beringer~\cite{Beringer11}.   
Our fourth alignment completeness theorem is for a logic including that rule together 
with lockstep rules but not \rn{SeqProd}.
As a worked example we show that $c2$ 
majorizes $c3$ 
for sufficiently large $x$.

Sec.~\ref{sec:discuss} discusses related work and open questions.  
Some recent works on relational verification 
use alignments that can be understood as more sophisticated product automata for which 
alignment complete rules remain to be designed.

\iflongver
\emph{This document extends the LICS'21 version with an appendix that contains additional details and proofs.}
\else
Additional details and proofs are in the appendix
of an extended version of the paper (\url{https://arxiv.org/abs/2101.11730}).
\fi

\section{Preliminaries and Floyd completeness}\label{sec:prelim}

In order to connect Floyd's theory with Hoare's we formulate the IAM 
in terms of transition systems with an explicit finite control flow graph (CFG).
We consider ordinary program syntax, with a standard structural operational semantics and Hoare logic, 
but with labels used to define the transition system of a given ``main program''.

\subsubsection*{Floyd automata, specs and correctness}

An \dt{automaton} is a tuple $(Ctrl,Sto,\init,\fin,\trans)$ where 
$Sto$ is a set (the data stores),
$Ctrl$ is a finite set (the control points)
that contains distinct elements $\init$ and $\fin$,
and ${\trans} \subseteq (Ctrl\times Sto)\times(Ctrl\times Sto)$ is the transition relation.
We require 
$(n,s)\trans (m,t)$ to imply $n\neq \fin$ and $n\neq m$ 
and call these the \dt{finality} and \dt{non-stuttering} conditions respectively.
Absence of stuttering loses no generality and facilitates definitions involving product automata.
Let $s,t$ range over stores and $n,m$ over control points.

A pair $(n,s)$ is called a \dt{state} and we write 
$\ctrl$, $\store$ for the left and right projections on states.
A \dt{trace} of an automaton is a non-empty sequence of states, consecutive under the transition relation.
A trace $\tau$ is \dt{terminated} provided $\tau$ is finite and $\ctrl(\last{\tau})=\fin$,
where $\last{\tau}$ denotes the last state of $\tau$. 
An \dt{initial trace} is one such that $\ctrl(\tau_0)=\init$.
We allow traces of length one, in which case $\last{\tau}=\tau_0$,
but a terminated initial trace has plural length because $\init\neq\fin$.  


We treat predicates semantically, i.e., as sets of stores. 
Define $s\models P$ iff $s\in P$ and define $(n,s)\models P$ iff $s\models P$.

Let us spell out two semantics for specs in terms of an automaton $A$.
The \dt{basic semantics} is as follows.
For a finite initial trace $\tau$ to satisfy $\spec{P}{Q}$ means that $\tau_0\models P$ 
and $\ctrl(\last{\tau})=\fin$ imply $\last{\tau}\models Q$, 
in which case we write $\tau\models \spec{P}{Q}$.
Then $A$ satisfies $\spec{P}{Q}$,
written $A \models \spec{P}{Q}$,
just if all its finite initial traces do.
For \dt{non-stuck semantics} there are two conditions:
(i) $\tau_0\models P$ and $\ctrl(\last{\tau})=\fin$ imply $\last{\tau}\models Q$, and 
(ii) $\tau_0\models P$ and $\ctrl(\last{\tau})\neq \fin$ imply $\last{\tau}\trans -$,
where $\last{\tau}\trans -$ means there is at least one successor state.
A state with no successor is called \dt{stuck}.
Non-final stuck states are often used to model runtime faults.
Again, $A$ satisfies $\spec{P}{Q}$ just if all its finite initial traces do.
Non-stuck is important in practice and we consider it in passing but for clarity 
our main development is for basic semantics.


For an automaton $A$, define $\CFG(A)$ to be the rooted directed graph with vertices
$Ctrl$, root $\init$, and an edge $n\cfge m$ iff $\some{s,t}{(n,s)\trans(m,t)}$.
For our purposes CFGs are unlabelled;
we write $n\cfge m$ for $(n,m)$ to avoid confusion with various other uses of pairs.
A \dt{path} is  a non-empty sequence of vertices 
that are consecutive under the edge relation.
By mapping the first projection ($\ctrl$) over a trace $\tau$ of $A$ we get its \dt{control path}, $\cpath(\tau)$, i.e., the sequence of control points in $\tau$.

A \dt{cutpoint set} for $A$ is a set $K\subseteq Ctrl$ with $\init\in K$ and 
$\fin\in K$, such that every cyclic path in $\CFG(A)$ contains at least one element of $K$.
Define $\segs(A,K)$, the \dt{segments} for $K$, to be the finite paths between cutpoints 
that have no intermediate cutpoint.
Formally, $vs\in \segs(A,K)$ iff
$vs$ is finite, 
$len(vs)>1$, 
$vs_0\in K$, 
$\last{vs}\in K$, and
$\all{i}{0<i<len(vs)-1 \imp vs_i\notin K}$.
A segment $vs$ is meant to refer to execution starting at control point $vs_0$ and 
ending at $\last{vs}$, hence the requirement $len(vs)>1$.
Note that $\segs(A,K)$ is finite because $\CFG(A)$ is finite and $K$ cuts every cycle.

\begin{figure*}[t]
\begin{center}
\begin{small}
\[ 
\begin{array}{ll}
c0: &
  \lassg{1}{y}{x};\lassg{2}{z}{1};
  \lwhilec{3}{y\neq 0}{ \lassg{4}{z}{z * y}; \lassg{5}{y}{y - 1}}
\\[.3ex]

c4: & 
\lassg{1}{y}{x}; \lassg{2}{z}{24}; \lassg{3}{w}{0}; 
   \lwhilec{4}{y\neq 4}{
      \lifc{5}{w \mod 2 = 0}{
        \lassg{6}{z}{z*y}; \lassg{7}{y}{y-1} }{\lskipc{8}}; \lassg{9}{w}{w+1} }
\\[.3ex]
c5: & 
\lassg{1}{y}{x}; \lassg{2}{z}{16}; \lassg{3}{w}{0}; 
    \lwhilec{4}{y\neq 4}{ 
          \lifc{5}{w \mod 3 = 0}{
           \lassg{6}{z}{z*2}; \lassg{7}{y}{y-1}}{\lskipc{8}}; \lassg{9}{w}{w+1} }
\end{array}
\]
\end{small}
\end{center}
\vspace*{-1ex}
\caption{Example labelled commands; $c4$ and $c5$ are variations on $c2$ and $c3$ of Fig.~\ref{fig:examples}.}\label{fig:c4c5}
\end{figure*}

\begin{figure}[t]
\begin{center}
\begin{footnotesize}

%
  \begin{tikzpicture}
  [->,every state/.style={fill=gray!10},
  initial text=$ $,
  auto,node distance=2.0cm,line width=0.1mm,inner sep=1pt,minimum size=10pt,
  scale=0.7,transform shape]
  
    \node[state, minimum size=1cm] (n1) {$1$};
    \node[state, right of=n1, minimum size=1cm] (n2) {$2$};
    \node[state, right of=n2, minimum size=1cm] (n3) {$3$};
    \node[state, right of=n3, minimum size=1cm] (n4) {$4$};
    \node[state, right of=n4, xshift=0.32cm, minimum size=1cm] (n5) {$5$};
    \node[state, right of=n5, minimum size=1cm] (n6) {$6$};

    \draw (n1) edge node[above]{$\lassg{}{y}{x}$} (n2)
          (n2) edge node[above]{$\lassg{}{z}{1}$} (n3)
          (n3) edge node[above]{$y\neq 0$} (n4)
          (n3) edge[above, bend left=40]
                    node[below]{$y=0$} (n6)
          (n4) edge node[above]{$\lassg{}{z}{z*y}$} (n5)
          (n5) edge [below, bend left=40]
                    node[above,yshift=0.1cm]{$\lassg{}{y}{y-1}$} (n3)
    ;
  \end{tikzpicture}
\end{footnotesize}
\end{center}
\vspace*{-2ex}
\caption{CFG of the automaton $\aut(c0;\lskipc{6})$, with suggestive edge labels.}\label{fig:c0aut}
\end{figure}
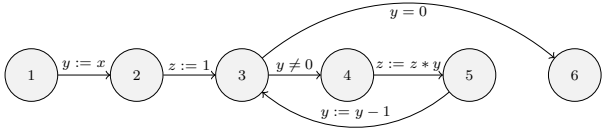

As an example, Fig.~\ref{fig:c4c5} shows a labelled version of program $c0$.
Fig.~\ref{fig:c0aut} shows the CFG of the automaton for
$c0;\lskipc{6}$.  The trailing skip serves to provide an end label.
The figure shows code as edge labels, for clarity,
but we do not formally consider edge-labelled CFGs.
One cutpoint set is $\{1,3,6\}$; its segments are
$[1,2,3]$, $[3,4,5,3]$, and $[3,6]$.

It is convenient to have notation for the effect of transitions along a segment.  
Given $vs\in\segs(A,K)$ define relation $\tranSeg{vs}$ by
$(n,s)\tranSeg{vs}(m,t)$ iff there is a trace $\tau$ of $A$ 
with $\cpath(\tau) = vs$ 
and $\tau_0=(n,s)$ and $\last{\tau}=(m,t)$.
Notice that $\tranSeg{vs}$ need not be total,
even in the typical case that the underlying relation $\trans$ is never stuck on non-$\fin$ states.
For example, if $vs_0$ represents a conditional branch
and in state $(vs_0,s)$ the condition does not drive the automaton to $vs_1$
then $(vs_0,s)$ is not in the domain of $\tranSeg{vs}$.

Given automaton $A$, cutpoint set $K$, and spec $\spec{P}{Q}$,
an \dt{annotation} is a function $an$ from $K$ to store predicates 
such that $an(\init)=P$ and $an(\fin)=Q$.
The requirement $\init\neq\fin$ ensures that annotations exist for any spec.

We lift $an$ to a function $\hat{an}$ that yields states:
\( \hat{an}(n) = \{ (n,s) \mid s\models an(n) \} \).
Put differently: $\hat{an}(n) = \{n\} \times an(n)$.
For each $vs$ in $\segs(A,K)$ there is a \dt{verification condition} (\dt{VC}):
\begin{equation}\label{eq:VC}
\POST(\tranSeg{vs})(\hat{an}(vs_0)) \subseteq \hat{an}(\last{vs}) 
\end{equation}
Here $\POST(\tranSeg{vs})$ is the direct image 
i.e., strongest postcondition.
Using the universal pre-image,\footnote{For relation $R$ 
   on states and set $X$ of states,
   $\PRE(R)(X) = \{ \alpha \mid \all{\beta}{\alpha R \beta \imp \beta\in X}\}$
   and $\POST(R)(X) = \{ \beta \mid \some{\alpha}{\alpha\in X \land \alpha R \beta } \}$.
   } 
(\ref{eq:VC}) is equivalent to
\( \hat{an}(vs_0) \subseteq \PRE(\tranSeg{vs})(\hat{an}(\last{vs})) \).
The VC says that for every trace $\tau$ with $\cpath(\tau)=vs$,
if \(\tau_0\models an(vs_0)\) then \(\last{\tau}\models an(\last{vs})\).
Annotation $an$ is \dt{valid} if all the VCs are true.
A segment represents a finite execution that follows that control path, 
so pre-image is the weakest precondition.

\begin{proposition}[soundness of IAM]\label{prop:iamsound}
Consider a valid annotation, $an$, of automaton $A$ with cutpoint set $K$, for $\spec{P}{Q}$.
In any initial trace $\tau$ of $A$
such that $\tau_0\models P$, at any position $i$, $0\leq i < len(\tau)$,
we have $\ctrl(\tau_i)\in K \imp \tau_i\models an(\ctrl(\tau_i))$.
Moreover $A\models \spec{P}{Q}$.
\end{proposition}

\begin{proposition}[completeness of IAM]\label{prop:iamcomplete}
Suppose $A\models \spec{P}{Q}$ and let $K$ be a cutpoint set.
Then there is an annotation on $K$ that is valid.
\end{proposition}

A \dt{full annotation} of an automaton is one where the cutpoint set is all control points.
Using strongest postconditions, including disjunction at control joins, one can show:

\begin{lemma}\label{lem:extendanno}
Any valid annotation can be extended to a full annotation that is valid.
\end{lemma}

In fact the extension can be constructed efficiently,
for many assertion languages and programming languages.  

\subsubsection*{Labelled commands}

A few tedious but routine technical details need to be spelled out 
in order to precisely formulate the main results.
Hoare logic is about programs;  
to make connections with automata we use syntax with labels $n\in\Z$:
\[ c::= 
\begin{array}[t]{l}
\lskipc{n} \mid \lassg{n}{x}{e}
  \mid c;c \mid \lchoice{n}{c}{c} \\
  \mid \lifc{n}{e}{c}{c} \mid \lwhilec{n}{e}{c}
\end{array}
\]
Here metavariable $x$ ranges over a set $\Var$ of variable names and
$e$ ranges over integer expressions.
The form $\lchoice{n}{c}{d}$ is for nondeterministic choice.
We also use metavariables $b,d,c',\ldots$ for commands.

Let $\lab(c)$ be the label of command $c$, defined recursively in the case of sequence:
$\lab(c;d) = \lab(c)$.
Let $\labs(c)$ be the set of labels that occur in  $c$.
The label of a command can be understood as its entry point.
We focus on programs of the form $c;\lskipc{\fin}$
where $\fin\in\nat$ serves as the exit label for $c$.  
Such a program can take at least one step,
even if $c$ is just skip; this fits with our formulation of automata.

Negative labels are used in the transition semantics, but for most purposes 
we are concerned with ``main programs'' which are required to have unique, non-negative labels.
This is formalized by the predicate $\ok(c)$ defined straightforwardly.
Fig.~\ref{fig:c4c5} gives example labelled commands.
The transition semantics is standard except for the manipulation of labels,
which is done in a way that facilitates definitions to come later.

As in the discussion of automata, let $s$ and $t$ range over stores ---but here we use 
\dt{variable stores}, i.e., total mappings from $\Var$ to $\Z$.
We write $\update{s}{x}{i}$ for the store like $s$ but mapping $x$ to $i$.
A \dt{configuration} $\config{c}{s}$ pairs a labelled command with a store,
and we let 
$\ctrl\config{c}{s}=c$ and $\store\config{c}{s}=s$.

\begin{figure}[t] 
\begin{small}
\begin{mathpar}
\inferrule{ s(e) \neq 0 }
{ \config{\lifc{n}{e}{c}{d}}{s} \ctrans \config{c}{s} }

\inferrule{ s(e) = 0 }
{ \config{\lifc{n}{e}{c}{d}}{s} \ctrans \config{d}{s} }

\inferrule{s(e) \neq 0 }
{ \config{\lwhilec{n}{e}{c}}{s} \ctrans 
  \config{c;\lwhilec{n}{e}{c}}{s}
}

\inferrule{s(e) = 0 }
{ \config{\lwhilec{n}{e}{c}}{s} \ctrans 
  \config{\lskipc{-n}}{s}
}

\inferrule{
\config{c}{s} \ctrans \config{d}{t} }
{ \config{c;b}{s} \ctrans \config{d;b}{t} }

\inferrule{}{ \config{\lassg{n}{x}{e}}{s} \ctrans \config{\lskipc{-n}}{\update{s}{x}{s(e)}} }

\inferrule{}{ \config{\lskipc{n};c}{s} \ctrans \config{c}{s} }

\inferrule{}{ \config{\lchoice{n}{c}{d}}{s} \ctrans \config{c}{s} }

\inferrule{}{ \config{\lchoice{n}{c}{d}}{s} \ctrans \config{d}{s} }

\end{mathpar}
\end{small}
\vspace*{-4ex}
\caption{Command semantics (with $n$ ranging over $\Z$).}\label{fig:progtrans}
\end{figure}

The transition relation $\ctrans$ is defined in Fig.~\ref{fig:progtrans}.
In a configuration reached from an $\ok$ command, 
the only negative labels are those introduced by the transitions for assignment and while,
which introduce negative labels on skip commands.
For while, one transition rule duplicates the loop body, creating non-unique labels.
For every $c,s$, either $\config{c}{s}$ has a successor or $c$ is $\lskipc{n}$ for some $n\in\Z$.
Assume integer expressions are everywhere defined, so configurations
are not stuck under $\ctrans$ unless the program is a single skip.



\subsubsection*{The automaton of a program}


If $\ok(c)$ and $m\in \labs(c)$, let $\sub(m,c)$ be the sub-command of $c$ with label $m$.
For example, consider $c0$ in Fig.~\ref{fig:c4c5},
then $\sub(2,c0)$ is $\lassg{2}{z}{1}$ and $\sub(3,c0)$ is
$\lwhilec{3}{y\neq 0}{\ldots}$.
%
%
To manipulate the CFG of a program of the form $c;\lskipc{\fin}$
that is $\ok$ (so, $\fin\notin\labs(c)$), we define functions $\fsuc$ and $\elab$.
For motivation, 
the control flow successors of the loop, \(sub(3,c0)\), in $c0;\lskipc{6}$ are 3 and 6.
Whereas 3 is inside $c0$, 6 is not.  
We call 6 the \dt{following successor},
given by $\fsuc(3,c0,6)$.
In general, $\fsuc(n,c,f)$ is defined by recursion on $c$; see Fig.~\ref{fig:fsuc}.
For example, $\fsuc(\sub(3,c0),c0,6)=6$
and $\fsuc(\sub(5,c0),c0,6)=3$.
For another example, let $c$ be
$\lifc{1}{x>0}{\lassg{2}{x}{x-1};\lassg{3}{y}{x}}{\lskipc{4}}$;
then 
$\fsuc(2,c,5)=3$ and
$\fsuc(1,c,5)= \fsuc(3,c,5) = \fsuc(4,c,5) =5$.

For subcommand $b$ of $c$ we define $\elab(b,c,\fin)$ to be the exit label,
i.e., the label to which control goes after every path through $b$.
In case $b$ is a conditional, loop, choice, 
assignment or skip, let $\elab(b,c,\fin) = \fsuc(\lab(b),c,\fin)$.
In case $b$ is a sequence $b_0;b_1$, let $\elab(b,c,\fin) = \elab(b_1,c,\fin)$,
i.e., the exit of a sequence is the exit of its last command.\footnote{Formally the definition of $\elab$ is by recursion on its first argument. We can define $\elab$ as a function of $b$ because unique labels
rules out multiple occurrences of a subprogram.
And it needs to be a function of command $b$, not its label, to handle sequences.}

\begin{figure}[t]
\begin{footnotesize}
\[
\begin{array}{l@{\;}c@{\;}l}
\fsuc(n,\lskipc{n},f)          & = & f \\
\fsuc(n,\lassg{n}{x}{e},f)     & = & f \\
\fsuc(n,c;d,f)                 & = & \fsuc(n,c,\lab(d)) \mbox{ , if $n\in\labs(c)$} \\
                               & = & \fsuc(n,d,f) \mbox{ , otherwise} \\ 
\fsuc(m, \lwhilec{n}{e}{c}, f) & = & f \mbox{ , if $m=n$} \\
                               & = & \fsuc(m,c,n) \mbox{ , otherwise} \\
\fsuc(m,\lifc{n}{e}{c}{d}, f) &= & \fsuc(m,c,f) \mbox{ , if $m\in \labs(c)$ }\\
                          & = & \fsuc(m,d,f) \mbox{ , if $m\in \labs(d)$ }\\
                          & = & f  \mbox{ , otherwise (i.e., $m=n$)} 
\\
\fsuc(m,\lchoice{n}{c}{d}, f) & = & \fsuc(m,c,f) \mbox{ , if $m\in \labs(c)$ }\\
                          & = & \fsuc(m,d,f) \mbox{ , if $m\in \labs(d)$ }\\
                          & = & f  \mbox{ , otherwise (i.e., $m=n$)} 
\end{array}
\]
\end{footnotesize}
\vspace*{-2ex}
\caption{Following successor $\fsuc(n,c,f)$, assuming 
$\ok(c)$, $n\in\labs(c)$, and $f\in\nat\setminus\labs(c)$.
}\label{fig:fsuc}
\end{figure}

Now we can define the CFG for an $\ok$ program $c;\lskipc{\fin}$, 
and with this in mind define an automaton with the same CFG to represent the program.
The nodes of the CFG are the control points $\{\fin\} \union \labs(c)$.
There is no control flow successor of $\fin$.
For $n\in\labs(c)$ there are one or two successors,
described by cases on $\sub(n,c)$:

\begin{small}
\begin{list}{}{}\item
\begin{tabular}{ll}
$\sub(n,c)$ & successor(s) of $n$ in the CFG \\\hline
$\lassg{n}{x}{e}$ & $n\cfge \fsuc(n,c,\fin)$ \\
$\lskipc{n}$ & $n\cfge \fsuc(n,c,\fin)$ \\
$\lifc{n}{e}{d_0}{d_1}$ & $n\cfge \lab(d_0)$ and $n\cfge \lab(d_1)$ \\
$\lchoice{n}{d_0}{d_1}$ & $n\cfge \lab(d_0)$ and $n\cfge \lab(d_1)$ \\
$\lwhilec{n}{e}{d}$ & $n\cfge\lab(d)$ and $n\cfge\fsuc(n,c,\fin)$
\end{tabular}
\end{list}
\end{small}



\noindent The automaton of an $\ok$ program $c;\lskipc{\fin}$,
written $\aut(c;\lskipc{\fin})$, is
\( (\labs(c)\union \{ \fin \}, (\Var\to\Z), \lab(c), \fin, \trans) \) 
where 
$(n,s)\trans (m,t)$ iff either
\begin{small}
\begin{ditemize}
\item $
\some{d}{ \config{\sub(n,c)}{s}\ctrans\config{d}{t} 
   \land \lab(d)\geq 0 \land m = \lab(d) }$
\item 
$\some{d}{ \config{\sub(n,c)}{s}\ctrans\config{d}{t} \land \lab(d) < 0 \land m = \fsuc(n,c,\fin) }$
\item or $
\sub(n,c)=\lskipc{n} \land m = \fsuc(n,c,\fin) \land t = s
$
\end{ditemize}
\end{small}
The first two cases use the semantics of Fig.~\ref{fig:progtrans}
for a sub-command on its own.  
The second case uses $\fsuc$ for a sub-command that has terminated.
The third case handles skip which on its own would be stuck, 
but which should take a step when it occurs as part of a sequence.
The only stuck states of $\aut(c;\lskipc{\fin})$ are terminated ones.



For any traces $\tau$ via $\ctrans$ and $\upsilon$ via $\trans$,
define $\tau\traceSim\upsilon$ iff
$len(\tau)=len(\upsilon)$ and 
$\store(\tau_i)=\store(\upsilon_i)$ and
$\lab(\ctrl(\tau_i)) = \ctrl(\upsilon_i)$, 
for all $i$, $0\leq i < len(\tau)$.

\begin{lemma}\label{lem:traceEq}
Suppose $\ok(c;\lskipc{n})$. Let $A$ be $\aut(c;\lskipc{n})$ and let $s$ be a store.
\begin{list}{}{}
\item[(i)]
For any trace $\tau$ from $\config{c;\lskipc{n}}{s}$ via $\ctrans$,
there is a trace $\upsilon$ of $A$ from $(\lab(c),s)$ via $\trans$,
such that $\tau\traceSim\upsilon$.
\item[(ii)]
For any trace $\upsilon$ of $A$ from $(\lab(c),s)$ via $\trans$, 
there is a trace $\tau$ from $\config{c;\lskipc{n}}{s}$ via $\ctrans$,
such that $\tau\traceSim\upsilon$.
\end{list}
\end{lemma}

For a full annotation, the segments are exactly the paths of length two,
i.e., the edges of the CFG.  This enables a straightforward
description of the VCs for the automaton of a program (not unlike the VCs given
by Floyd~\cite{Floyd67} for flowchart programs).

\begin{figure*}
\begin{center}
\begin{small}
\begin{tabular}{lll}
if $b = \sub(n,c)$ is\ldots & and $n\cfge m$ in CFG is\ldots & then the VC is equivalent to\ldots \\\hline

$\lskipc{n}$ &
$m=\elab(b,c,\fin)$ & $an(n)\imp an(m)$ 
\\

$\lassg{n}{x}{e}$ &
$m=\elab(b,c,\fin)$ & $an(n)\imp \subst{an(m)}{x}{e}$ 
\\

$\lifc{n}{e}{d_0}{d_1}$ & $m=\lab(d_0)$ & $an(n)\land e \imp an(m)$ 
\\

$\lifc{n}{e}{d_0}{d_1}$ & $m=\lab(d_1)$ & $an(n)\land \neg e \imp an(m)$ 
\\

$\lwhilec{n}{e}{d}$ & $m=\lab(d)$&  $an(n)\land e \imp an(m)$ 
\\

$\lwhilec{n}{e}{d}$ & $m=\elab(b,c,\fin)$ & $an(n)\land \neg e \imp an(m)$
\\

$\lchoice{n}{d_0}{d_1}$ & $m$ is $\lab(d_0)$ or $\lab(d_1)$ & $an(n) \imp an(m)$ 
\end{tabular}

\end{small}
\end{center}
\vspace*{-1ex}
\caption{VCs for the automaton of a program $c;\lskipc{\fin}$ and full annotation $an$.}\label{tab:VC}
\end{figure*}

\begin{lemma}[VCs for programs]\label{lem:VCprog}
Consider an $\ok$ program $c;\lskipc{\fin}$ and 
a full annotation, $an$, of its automaton.
For each control edge $n\cfge m$, the VC (\ref{eq:VC}) 
can be expressed as in Fig.~\ref{tab:VC}.
\end{lemma}
The conditions 
are derived straightforwardly from the semantic definitions.
Although we are treating assertions as sets of stores, 
we use formula notations like $\land$ and $\imp$, rather than $\intersect$ and $\subseteq$,
for clarity.
We write $an(n)\land e$ to abbreviate $an(n)\intersect\{s \mid s(e)\neq 0\}$.
For a set $P$ of program stores, we use substitution notation $\subst{P}{x}{e}$
with standard meaning: $s\in\subst{P}{x}{e}$ iff $\update{s}{x}{s(e)}\in P$.

\subsubsection*{Floyd completeness of Hoare Logic}

\begin{figure}[t]
\begin{small}
\begin{mathpar}
\mprset{sep=1.3em}
\inferrule[Skip]{}{ 
\skipc:\spec{P}{P} 
}

\inferrule[Ass]{}{ 
x:=e : \spec{\subst{P}{x}{e}}{P}
}

\inferrule[Seq]
{
c:\spec{P}{R} \\ d:\spec{R}{Q}
}{
c;d : \spec{P}{Q} 
}

\inferrule[If]
{
c:\spec{P\land e}{Q} \\ d:\spec{P\land \neg e}{Q}
}{
\ifc{e}{c}{d} : \spec{P}{Q} 
}

\inferrule[Wh]
{
c:\spec{P\land e}{P} 
}{
\whilec{e}{c} : \spec{P}{P\land \neg e} 
}

\inferrule[Choice]
{
c:\spec{P}{Q} \\ d:\spec{P}{Q}
}{
\choice{c}{d} : \spec{P}{Q} 
}

\inferrule*[left=Conseq\quad]
{
P\imp R \\ 
c : \spec{R}{S} \\
S\imp Q
}{
c : \spec{P}{Q} 
}
\end{mathpar}
\end{small}
\vspace*{-4ex}
\caption{Rules of HL (command labels elided).}\label{fig:HL}
\end{figure}

Fig.~\ref{fig:HL} gives the rules of HL.
We write $\proves$ to indicate derivability using the rules. 
As usual, the semantics is that for all $s,t$, if $s\models P$ and
$\config{c}{s}\ctrans^* \config{\lskipc{n}}{t}$
then $t\models Q$.
We write this as $\models c:\spec{P}{Q}$.  
As is well known, the rules are sound: $\proves c: \spec{P}{Q}$ implies $\models c: \spec{P}{Q}$.  

A corollary of Lemma~\ref{lem:traceEq} is that if $\ok(c;\lskipc{f})$ then 
$\aut(c;\lskipc{f})\models \spec{P}{Q}$
iff 
$\models c;\lskipc{f} : \spec{P}{Q}$.
The trailing skip loses no generality.
By semantics, 
$\models c;\lskipc{f} : \spec{P}{Q}$ iff $\models c: \spec{P}{Q}$.
In terms of proofs, the two are equi-derivable.

Given a valid annotation for $\aut(c;\lskipc{\fin})$ and $\spec{P}{Q}$,
by Prop.~\ref{prop:iamsound} we have $\aut(c;\lskipc{\fin})\models \spec{P}{Q}$,
so by the corollary of Lemma~\ref{lem:traceEq}
we have $\models c;\lskipc{\fin} : \spec{P}{Q}$
and thus $\models c: \spec{P}{Q}$.
So by the standard (Cook) completeness result for HL~\cite{Cook78,AptOld3}
there is a proof of $c : \spec{P}{Q}$.
The idea of Floyd completeness is that from a valid annotation $an$ one may 
obtain a proof that essentially uses only 
judgments given directly by the annotation. 
For a full annotation, $an$, of $\aut(c;\lskipc{\fin})$,
define the \dt{associated judgments} to be:
\begin{ditemize}
\item for subprograms $b$ of $c$, the judgments
\begin{equation}\label{eq:anno}
 b : \spec{ an(\lab(b)) }{ an(\elab(b,c,\fin)) }
\end{equation}
\item $ b : \spec{ an(\lab(b)) \land e}{ an(\elab(b,c,\fin)) }$
where $b$ is the body of a loop, or then-branch of a conditional, with test $e$;
\item $ b : \spec{ an(\lab(b)) \land \neg e}{ an(\elab(b,c,\fin)) }$ 
where $b$ is the else-branch of a conditional with test $e$;
\item $ b : \spec{ an(\lab(b)) }{ an(\lab(b)) \land \neg e } $
where $b$ is a loop with test $e$; and 
\item $ \lassg{n}{x}{e} : \spec{ \subst{an(m)}{x}{e} }{ an(m) } $
   where $m=\elab(\lassg{n}{x}{e},c,\fin)$.
\end{ditemize}

\begin{theorem}[Floyd completeness] \label{thm:floydComplete}
Consider an $\ok$ program $c;\lskipc{\fin}$,
and a valid annotation, $an$, of $\aut(c;\lskipc{\fin})$ for $\spec{P}{Q}$.
Then there is proof in HL of $c: \spec{P}{Q}$
using only the associated judgments of $an$.
\end{theorem}

A corollary is Cook completeness, i.e., 
$\models c: \spec{P}{Q}$ implies $\proves c: \spec{P}{Q}$,
using Prop.~\ref{prop:iamcomplete}.

To prove the theorem, we first prove that 
\begin{equation}\label{eq:claim}
\proves b : \spec{ an(\lab(b)) }{ an(\elab(b,c,\fin)) }
\end{equation}
for every subprogram $b$ of $c$.
The claim (\ref{eq:claim}) is proved by structural induction on $c$.
In each case, we use one instance of the syntax-directed rule for $b$, and in some cases also \rn{Conseq}.
The base cases are the assignments and skip commands in $c$.
For such a command $b$, let $m=\elab(b,c,\fin)$.
\begin{ditemize}

\item If $b$ is $\lskipc{n}$, 
we have
\ghostbox{$\proves \lskipc{n} : \spec{ an(m) }{ an(m) }$}
by rule \rn{Skip},
and Lemma~\ref{lem:VCprog} gives $an(n)\imp an(m)$
(using validity of $an$),
so by \rn{Conseq} we get 
\ghostbox{$\proves \lskipc{n} : \spec{ an(n) }{ an(m) }$}

\item if $b$ is $\lassg{n}{x}{e}$, 
we have 
\ghostbox{$\proves \lassg{n}{x}{e} : \spec{ \subst{an(m)}{x}{e} }{ an(m) }$} by rule \rn{Ass},
and Lemma~\ref{lem:VCprog} gives $an(n)\imp \subst{an(m)}{x}{e}$,  
so by \rn{Conseq} we get 
\ghostbox{$\proves \lassg{n}{x}{e} : \spec{ an(n) }{ an(m) }$} 
\end{ditemize}
The induction step is as follows.
(Other cases in appendix.) 
\begin{ditemize}
\item If $b$ is the sequence $b_0;b_1$, 
by induction we have
\ghostbox{$\proves b_0: \spec{an(\lab(b_0))}{an(\elab(b_0,c,\fin))}$} and 
\ghostbox{$\proves b_1: \spec{an(\lab(b_1))}{an(\elab(b_1,c,\fin)}$}.
We have
$\elab(b_0,c,\fin)=\lab(b_1)$ and
$\elab(b_0;b_1,c,\fin)=\elab(b_1,c,\fin)$,
so by \rn{Seq} we get 
\ghostbox{$\proves b : \spec{ an(\lab(b)) }{ an(\elab(b,c,\fin)) }$}.

\item If $b$ is $\lifc{n}{e}{d_0}{d_1}$,    by induction we have 
\ghostbox{$\proves d_0 : \spec{ an(\lab(d_0) }{ an(\elab(d_0,c,\fin)) }$}
and 
\ghostbox{$\proves d_1 : \spec{ an(\lab(d_1) }{ an(\elab(d_1,c,\fin)) }$}.
Lemma~\ref{lem:VCprog} gives
$an(n)\land e \imp an(\lab(d_0))$ and $an(n)\land \neg e \imp an(\lab(d_1))$  
so by Conseq we get 
\ghostbox{$\proves d_0 : \spec{ an(n)\land e }{ an(\elab(d_0,c,\fin)) }$}
and 
\ghostbox{$\proves d_1 : \spec{ an(n)\land \neg e }{ an(\elab(d_1,c,\fin)) }$}.
By definitions we have $\elab(b,c,\fin) = \elab(d_0,c,\fin) = \elab(d_1,c,\fin)$. 
So rule \rn{If} yields 
\ghostbox{$\proves b : \spec{ an(n) }{ an(\elab(b,c,\fin) }$}.

\end{ditemize}
To prove the theorem we instantiate (\ref{eq:claim}) with $c$ itself, to get
$\proves c : \spec{ an(\lab(c)) }{ an(\elab(c,c,\fin)) }$.
Now $\init= \lab(c)$ by definition of $\aut(c;\lskipc{\fin})$,
and $an$ is an annotation for $\spec{P}{Q}$,
so $an(\lab(c)) = an(\init) = P$ and $an(\elab(c,c,\fin)) = an(\fin) = Q$.
Thus we have obtained $\proves c : \spec{ P }{ Q }$,
highlighting the associated judgments, q.e.d.

In the rest of the paper, we assume without mention that all considered programs satisfy $\ok$.  

\section{Relational judgments and product automata}\label{sec:prodAut}

Let $A = (Ctrl,Sto,\init,\fin,\trans)$ and 
$A' = (Ctrl',Sto',\init',\fin',\trans')$ be automata.
A relational spec $\rspec{\R}{\S}$
is comprised of relations $\R$ and $\S$ from $Sto$ to $Sto'$.
We write  $s,s'\models\R$ iff $(s,s')\in\R$ and 
$(n,s),(n',s')\models \R$ iff  $s,s'\models\R$.
Finite traces $\tau$ of $A$ and $\tau'$ of $A'$ satisfy 
$\rspec{\R}{\S}$,
written $\tau,\tau'\models \rspec{\R}{\S}$,
just if $\tau_0,\tau'_0\models \R$, $ctrl(\last{\tau})= \fin$, and $ctrl(\last{\tau'})=\fin'$ imply
$\last{\tau},\last{\tau'}\models \S$.
The pair $A,A'$ satisfies $\rspec{\R}{\S}$,
written $A|A'\models\rspec{\R}{\S}$, just if all pairs of finite initial traces do.

In passing we will consider the \dt{non-stuck semantics of relational specs}:
in addition to the above conditions, it requires, for all finite initial $\tau,\tau'$ 
such that $\tau_0,\tau'_0\models\R$, 
that $ctrl(\last{\tau})\neq \fin$ implies $\last{\tau}\trans -$
and  $ctrl(\last{\tau'})\neq \fin'$ implies $\last{\tau'}\trans' -$.

In casual examples, we use primed identifiers in specs to refer to the second execution.
The sequential product rule involves renaming identifiers in order to encode
two computations as one, but our product constructions and RHL rules do
not require programs to act on distinct variables.  We continue to use primes on metavariables
to aid the reader, but introduce notations like $\eqbi{x}{x}$ which expresses
equality of the values of $x$ in two states with the same variables.
For program expressions, $\eqbi{e}{e'}$ denotes the relation 
$\{(s,s') \mid s(e) =  s'(e') \}$ which we call \dt{agreement}.
For example, $\rspec{\eqbi{x}{x}}{\eqbi{z}{z}}$ expresses that the final value of $z$ 
is determined by the initial value of $x$.
Owing to our use of non-zero integers to represent truth
(in semantics Fig.~\ref{fig:progtrans}),
we also need a different form, $\eqbib{e}{e'}$, to express agreement on truth value; it denotes
$\{(s,s') \mid s(e) \neq 0 \mbox{ iff } s'(e')\neq 0 \}$. 
We also write 
$\leftF{e}$ for the set $\{ (s,t) \mid s(e)\neq 0 \}$
and $\rightF{e}$ for $\{ (s,t) \mid t(e)\neq 0 \}$.

In the presence of nondeterminacy, the $\forall\forall$-properties 
expressed by specs $\rspec{\R}{\S}$ are not the only properties of interest.
For example, the program $\choice{x:=x+1}{x:=x+2}$ does not satisfy   
$\rspec{\eqbi{x}{x}}{\eqbi{x}{x}}$,
but it does satisfy the ``possibilistic noninterference'' property that for any two
stores $s,s'$ that agree on $x$, and any run from $s$, there exists a run from $s'$ 
with the same final value for $x$.  
Such $\forall\exists$-properties are beyond the scope of this paper.
The nondeterministic program 
$(\choice{y:=0}{y:=1});x:=x+1$ does satisfy $\rspec{\eqbi{x}{x}}{\eqbi{x}{x}}$.

A product of automata $A$ and $A'$ is 
meant to represent a chosen alignment 
of steps of $A$ with steps of $A'$.  
In many cases, the control points of a product are pairs $(n,m)$ from underlying automata,
but we continue to use identifiers $n,m$ for control points regardless of their type.

A \dt{product} of $A$ and $A'$ is an automaton $\Pi_{A,A'}$ of the form
\( (C, (Sto\times Sto'), i, f, \biTrans)
\), 
together with functions $\LT:C\to Ctrl$ and $\RT:C\to Ctrl'$
such that the following hold: 
First, $\LT(i) = \init$, $\RT(i)=\init'$, $\LT(f) = \fin$, and $\RT(f)=\fin'$.
Second, $ (n,(s,s')) \biTrans (m,(t,t')) $ implies either
   \begin{ditemize}
   \item  $(\LT(n),s)\trans(\LT(m),t)$ and  $(\RT(n),s') = (\RT(m),t')$, or
   \item  $(\LT(n),s) = (\LT(m),t)$ and  $(\RT(n),s')\trans'(\RT(m),t')$, or  
   \item  $(\LT(n),s)\trans(\LT(m),t)$ and  $(\RT(n),s')\trans'(\RT(m),t')$.
   \end{ditemize}
The second condition says each step of the product represents a step of $A$, a step of $A'$, or both.
For states of $\Pi_{A,A'}$, define 
$\Left(n,(s,s')) = (\LT(n),s)$ and $\Right(n,(s,s')) = (\RT(n),s')$.
We extend $\Left$ and $\Right$ to traces of $\Pi_{A,A'}$ by
$\Left(T) = \destutter(\mathconst{map}(\Left,T))$ and 
$\Right(T) = \destutter(\mathconst{map}(\Right,T))$.
Observe that $\Left(T)$ is a trace of $A$ and $\Right(T)$ a trace of $A'$.
(Any repeated states in $\mathconst{map}(\Left,T)$ are not transitions of $A$,
owing to the non-stuttering condition for automata.)

A product is \dt{$\R$-adequate} if it covers all terminated initial traces from $\R$-states:
For all terminated initial traces $\tau$ of $A$ and $\tau'$ of $A'$ 
such that $\tau_0,\tau'_0\models\R$, 
there is a trace $T$ of $\Pi_{A,A'}$ 
with $\tau = \Left(T)$ and $\tau' = \Right(T)$.
A product is \dt{adequate} if it is adequate for all initial pairs of states (i.e., true-adequate).
A product is \dt{strongly $\R$-adequate} if it covers prefixes of diverging traces, in addition to terminated ones, that is:
For all finite initial traces $\tau,\tau'$ of $A,A'$
such that $\tau_0,\tau'_0\models\R$, 
there is a trace $T$ of $\Pi_{A,A'}$ 
such that $\tau\prefixeq\Left(T)$ and $\tau'\prefixeq\Right(T)$,
where $\prefixeq$ means prefix.
Moreover, it does not have \dt{one-sided divergence}:
there is no infinite trace $T$, with $T_0\models\R$,
with $i$ such that $\Left(T_j)=\Left(T_i)$ for all $j>i$,
or $\Right(T_j)=\Right(T_i)$ for all $j>i$.
Note that strong $\R$-adequacy implies $\R$-adequacy.

Here are some products defined for arbitrary $A,A'$,
taking $C$ to be $Ctrl\times Ctrl'$ and $\LT,\RT$ to be $\fst,\snd$.

\begin{small}
\begin{list}{}{}
\item[\dt{only-lockstep}.] 
$((n,n'),(s,s')) \biTrans_{olck} ((m,m'),(t,t'))$ iff 
\\
$(n,s)\trans(m,t)$ and $(n',s')\trans'(m',t')$.

\item[\dt{left-only}.]
$((n,n'),(s,s')) \biTrans_{lo} ((m,m'),(t,t'))$ iff 
\\
$(n,s)\trans(m,t)$ and $(n',s')=(m',t')$.

\item[\dt{right-only}.]
$((n,n'),(s,s')) \biTrans_{ro} ((m,m'),(t,t'))$ iff 
\\
$(n,s)=(m,t)$ and $(n',s')\trans'(m',t')$.

\item[\dt{interleaved}.]
The union $\biTrans_{lo} \union \biTrans_{ro}$.

\item[\dt{eager-lockstep}.]
$((n,n'),(s,s')) \biTrans_{elck} ((m,m'),(t,t'))$ iff 
\\
$((n,n'),(s,s')) \biTrans_{olck} ((m,m'),(t,t'))$, or 
\\
$n=\fin$ and $((n,n'),(s,s')) \biTrans_{ro} ((m,m'),(t,t'))$, or
\\
$n'=\fin'$ and $((n,n'),(s,s')) \biTrans_{lo} ((m,m'),(t,t'))$



\item[\dt{sequential}.] 
$((n,n'),(s,s')) \biTrans_{seq} ((m,m'),(t,t'))$ iff  
\\
$n'=\init'$ and $((n,n'),(s,s')) \biTrans_{lo} ((m,m'),(t,t'))$, or 
\\
$n=\fin$ and $((n,n'),(s,s')) \biTrans_{ro} ((m,m'),(t,t'))$.


\item[\dt{ctrl-conditioned}.]
Given subsets $L,R,J$ of $Ctrl\times Ctrl'$, define
$((n,n'),(s,s')) \biTrans_{cnd} ((m,m'),(t,t'))$ iff  
\\
$(n,n')\in L$ and $((n,n'),(s,s')) \biTrans_{lo} ((m,m'),(t,t'))$, or 
\\
$(n,n')\in R$ and $((n,n'),(s,s')) \biTrans_{ro} ((m,m'),(t,t'))$, or 
\\
$(n,n')\in J$ and $((n,n'),(s,s')) \biTrans_{olck} ((m,m'),(t,t'))$.
\end{list}
\end{small}
As an example of the ctrl-conditioned form,
the \dt{lockstep-control} product restricts only-lockstep to additionally 
require the two executions to follow the same control path.
(So it only reaches a linear number out of the quadratically many control points.)
This can be described by taking $L=R=\emptyset$ and defining the condition for joint 
steps by $(n,n')\in J$ iff $n=n'$.
The reader can check that the ctrl-conditioned form subsumes the others listed above. 

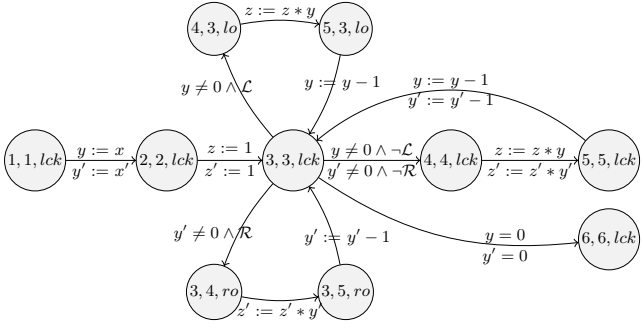
\begin{figure}
\begin{small}
\begin{center}
  \begin{tikzpicture}
    [->,every state/.append style={fill=gray!10},
initial text=$ $,
auto,node distance=2.5cm,line width=0.1mm,inner sep=1pt,
scale=0.70,transform shape]
  \node[state] (lck1) {$1,1,lck$};
  \node[state, right of=lck1] (lck2) {$2,2,lck$};
  \node[state, right of=lck2, xshift=-0.1cm] (lck3) {$3,3,lck$};
  \node[state, right of=lck3, xshift=0.5cm] (lck4) {$4,4,lck$};
  \node[state, right of=lck4, xshift=0.5cm] (lck5) {$5,5,lck$};
  \node[state, below of=lck5, yshift=1cm] (lck6) {$6,6,lck$};
  \node[state, above of=lck3, xshift=-1.5cm] (lo4) {$4,3,lo$};
  \node[state, right of=lo4] (lo5) {$5,3,lo$};
  \node[state, below of=lck3, xshift=-1.5cm] (ro4) {$3,4,ro$};
  \node[state, right of=ro4] (ro5) {$3,5,ro$};
  \draw
  (lck1) edge node[above]{$\lassg{}{y}{x}$}
              node[below]{$\lassg{}{y'}{x'}$} (lck2)
  
  (lck2) edge node[above,yshift=0.1cm]{$\lassg{}{z}{1}$}
              node[below]{$\lassg{}{z'}{1}$} (lck3)

 

  (lck3) edge node[above]{$y\neq 0 \land \neg\Lrel$}
              node[below]{$y'\neq 0 \land \neg\R$} (lck4)
  
  (lck4) edge node[above]{$\lassg{}{z}{z*y}$} node[below]{$\lassg{}{z'}{z'*y'}$} (lck5)

  (lck3) edge[bend left=10] node[left,yshift=0.2cm,xshift=0.2cm]{$y\neq0 \land \Lrel$} (lo4)

  (lo4) edge[bend left=10] node[above]{$\lassg{}{z}{z*y}$} (lo5)

  (lck3) edge[bend right=10] node[left,yshift=-0.2cm,xshift=0.2cm]{$y'\neq0 \land \R$} (ro4)

  (ro4) edge[bend right=10] node[below]{$\lassg{}{z'}{z'*y'}$} (ro5)

  (lo5) edge[bend left=10] node[right,xshift=-0.5cm,yshift=0.25cm]{$\lassg{}{y}{y-1}$} (lck3)

  (ro5) edge[bend right=10] node[right,xshift=-0.5cm,yshift=-0.25cm]{$\lassg{}{y'}{y'-1}$} (lck3)

  (lck3) edge[bend right=20] node[above,xshift=1.2cm,yshift=-0.18cm]{$y=0$}
                            node[below,xshift=1.2cm,yshift=-0.18cm]{$y'=0$} (lck6)
  (lck5) edge[bend right=40]  node[above]{$\lassg{}{y}{y-1}$}
                              node[below]{$\lassg{}{y'}{y'-1}$} (lck3)
  ;
  \end{tikzpicture}
\end{center}
\end{small}
\vspace*{-1ex}
\caption{
Conditionally aligned loop product automaton for $c0\sep c0$,
with informal edge labels including alignment guards $\Lrel,\R$.}\label{fig:aut3}
\end{figure}

A product can also be conditioned on data, as explored in Sec.~\ref{sec:dissonant}.
Fig.~\ref{fig:aut3} depicts a product of $c0$ with itself,
in which an iteration of the loop body on just the left (resp.\ right) side
may happen only when the store relation $\Lrel$ (resp.\ $\R$) holds.

The only-lockstep form is not adequate, in general, because a terminated state can be reached on one side before the other side terminates.
But even lockstep-control can be $\R$-adequate in some cases, 
for example let $\R$ be equality of stores and $A=A'$,
then lockstep-control is $\R$-adequate if $A$ is deterministic.

The interleaved 
product is strongly adequate---but not very helpful, 
since so many pairs of control points are reachable.
The sequential form is adequate but not strongly adequate:
if $\tau$ is a finite prefix of a divergent trace, and $\tau'$ has length $>1$,
the sequential product never finishes on the left and so does not cover $\tau'$.

Eager-lockstep is adequate, and strongly adequate if 
the underlying automata have no stuck states.
It shows the need for prefix in the definition of strong adequacy:
if $\tau$ is shorter than $\tau'$, the product automaton  needs to extend $\tau$ 
by further steps in order to cover $\tau'$.

Auxiliary control state can be used to ensure strong adequacy.
The \dt{dovetail product}
has $C = Ctrl\times Ctrl' \times \{0,1\}$ with
$\LT(n,n',i)=n$ and $\RT(n,n',i)=n'$.
Let $ ((n,n',i),(s,s')) \biTrans_{dov} ((m,m',j),(t,t')) $ iff 
either 
\begin{itemize}
\item $i=0$, $j=1$, $((n,n')(s,s'))\biTrans_{lo}((m,m'),(t,t'))$, or 
\item $i=1$, $j=0$, $((n,n')(s,s'))\biTrans_{ro}((m,m'),(t,t'))$, or 
\item one of the last two eager-lockstep cases apply (one side terminated).
\end{itemize}

The most general notion of product allows auxiliary store in addition to auxiliary control.
We return to this in Sec.~\ref{sec:discuss}.

Owing to the definition of stores of a product, 
a relation $\R\subseteq Sto\times Sto'$ from stores of $A$ to stores of $A'$
is the same thing as a predicate on stores of a product $\Pi_{A,A'}$.
So a relational spec $\rspec{\R}{\S}$ for $A,A'$ can be seen as a
unary spec $\spec{\R}{\S}$ for $\Pi_{A,A'}$.

\begin{theorem}\label{thm:product}
For the basic semantics of specs,
if $\Pi_{A,A'}$ is an $\R$-adequate product of $A,A'$
then $\Pi_{A,A'} \models \spec{\R}{\S}$ iff 
$A|A' \models \rspec{\R}{\S}$.
For the non-stuck semantics, 
if $\Pi_{A,A'}$ is a strongly $\R$-adequate  product 
then $\Pi_{A,A'} \models \spec{\R}{\S}$ implies $A|A' \models \rspec{\R}{\S}$.
\end{theorem}
This lifts IAM to a method for proving a relational judgment for programs:
construct their automata, define a product $\Pi$ and an annotation $an$; prove $\R$-adequacy of $\Pi$ and validity of $an$.

The number of cutpoints needed for a product may be on the order of the product of the number 
for the underlying automata.
But some products have fewer.  
Unreachable cutpoints can be annotated as false
so the corresponding VCs are vacuous.

\section{A one-rule complete RHL}\label{sec:seqcomplete}

Recall the sequential product rule sketched in Sec.~\ref{sec:intro}.
Sequential product is adequate so by Theorem~\ref{thm:product} 
it can be used to prove correctness for any relational spec.
This leads to the well known rule that we now study in detail.

Although the definition of automaton allows an arbitrary set for stores, 
in HL we require stores to be mappings on variables, specifically on the set $\Var$. 
To express a product as a single program we need to encode a pair of such stores as a single one.
Let $\Dot{\Var}$ be the set of fresh variable names 
$\Dot{x}$ such that $x\in\Var$.  Let 
$\fdot:\Var\to\Dot{\Var}$ be the obvious bijection.
Given $s:\Var\to\Z$ and $t:\Dot{\Var}\to\Z$, the union $s\union t$ is a function $\Var\union\Dot{\Var}\to\Z$
(treating functions as sets of ordered pairs).  
For $s$ and $t$ of type $\Var\to\Z$ define $s+t = s\union(t\circ \fdot^{-1})$,
so $s+t : \Var\union\Dot{\Var}\to\Z$ faithfully represents $(s,t)$.
For relation $\R$ on variable stores, let $\R^\eplus$ be the predicate on $\Var\union\Dot{\Var}\to\Z$
given by $ \R^\eplus = \{ s+t \mid (s,t)\in \R \}$.
We overload the name $\fdot$ for the function renaming variables of an expression,
and for command $c$ on $\Var$ let $\fdot(c)$ be the command on $\Dot{\Var}$ obtained by renaming;
this leaves labels unchanged.


Define semantic substitution $\subst{\R}{x|x'}{e|e'}$ by  
$(s,t)\in\subst{\R}{x|x'}{e|e'}$ iff $(\update{s}{x}{s(e)},\update{t}{x'}{t(e')})\in \R$.
(Here $x$ need not be distinct from $x'$.)
Substitution is preserved by the encoding: 
\begin{equation}\label{eq:substcommute}
(\subst{\R}{x|x'}{e|e'})^\eplus = \subst{(\R^\eplus)}{x,\Dot{x'}}{e,\Dot{e'}}
\end{equation}
(Using simultaneous substitution on the right; it can as well be written
$\subst{(\subst{(\R^\eplus)}{x}{e})}{\Dot{x'}}{\Dot{e'}}$.)
We write $\subst{\R}{x|}{e|}$ for substitution that leaves the right side unchanged.
We refrain explicit notation to distinguish between correctness judgments for $\Var$- programs and those for $\Var\union\Dot{\Var}$-programs.

\begin{lemma}
\label{lem:product}
Let $c,d$ be programs on $\Var$ and $\R,\S$ be relations on $\Var$-stores.
Let $d'=\fdot(d)$, so  $c;d'$ is a $(\Var\union\Dot{\Var})$-program.
Then $\models c;d' : \spec{\R^\eplus}{\S^\eplus}$ iff $\models c|d : \rspec{\R}{\S}$.
\end{lemma}

The lemma implies soundness of the following sequential product rule
(sometimes called self composition):
\begin{mathpar}
\inferrule*[left=SeqProd\quad]
{ d' = \fdot(d) \\ c;d' : \spec{\R^\eplus}{\S^\eplus} }
{ c\sep d : \rspec{\R}{\S} }
\end{mathpar}

\begin{proposition}[one-rule complete RHL]
\label{prop:RHLoneComplete}
The logic comprised of \rn{SeqProd} and the rules of HL (Fig.~\ref{fig:HL}) is complete.
\end{proposition}
To prove this, suppose $\models c|d : \rspec{\R}{\S}$.
By Lemma~\ref{lem:product} we have 
$\models c;d' :  \spec{\R^\eplus}{\S^\eplus}$ where $d'=\fdot(d)$.
By completeness of HL we have 
$\proves c;d' : \spec{\R^\eplus}{\S^\eplus}$
so \rn{SeqProd} yields 
$\proves c|d : \rspec{\R}{\S}$, q.e.d.

Analogous to Floyd completeness (Thm.~\ref{thm:floydComplete}),
we can directly prove a stronger result.  
It refers to the associated judgments of an annotation of a sequential product.
These are similar to those defined preceding Thm.~\ref{thm:floydComplete},
but now they have the form
$b: \spec{an(n,1)^\eplus}{an(m,1)^\eplus}$ (with $b$ a subprogram of $c$ with $n=\lab(b)$ and $m$ the end label),
the form 
$b: \spec{an(\fin,n)^\eplus}{an(\fin,m)^\eplus}$,
and the variations like preceding Thm.~\ref{thm:floydComplete}.
The exact definition of associated judgment becomes evident in the proof to follow.

\begin{theorem}[alignment completeness for sequential product]\label{thm:seqcomplete}
If $an$ is a valid full annotation of the sequential product of $\aut(c;\lskipc{\fin})$ and $\aut(d;\lskipc{\fin})$, 
for spec $\spec{\R}{\S}$, 
then $\proves c\sep d : \rspec{\R}{\S}$
in the logic comprised of HL and \rn{SeqProd}.
Moreover this judgment can be proved using only the 
associated judgments. 
\end{theorem}
As a corollary, any valid annotation of a sequential product gives a provable judgment,
using loop invariants given by the annotation,
because the annotation can be extended to a full one (Lemma~\ref{lem:extendanno}).
We choose to state all the theorems for full annotations, just to avoid a more 
complicated definition for the associated judgments.

The proof relies on an analysis like Lemma~\ref{lem:VCprog} 
but for VCs of sequential product.
Assume w.l.o.g.\ that $\lab(c)=1=\lab(d)$.
As a first step, consider the CFG of the sequential product $\Pi$ of $\aut(c;\lskipc{\fin})$ and $\aut(d;\lskipc{\fin})$.
Initial states of $\Pi$ have the form $((1,1),(s,t))$,
final states have the form $((\fin,\fin),(s,t))$, and edges of the CFG are of two forms:
$(n,1)\cfge (m,1)$ for $n\cfge m$ in the CFG of $c;\lskipc{\fin}$ and 
$(\fin,n)\cfge(\fin,m)$ for $n\cfge m$ in the CFG of $d;\lskipc{\fin}$.

Now assume $an$ is a full annotation of $\Pi$.  Because steps of $\Pi$ represent execution of a program
on one side or the other, the VCs are similar to those in Fig.~\ref{tab:VC} except that they
pertain to control points of the forms $(n,1)$ and $(\fin,n)$.
To save space we just give some illustrative cases in Fig.~\ref{tab:VCseq},
using some notations from Sec.~\ref{sec:prodAut}.

\begin{figure*}
\begin{center}
\begin{small}
\begin{tabular}{lll}
if $b = \sub(n,c)$ is\ldots & and $(n,1)\cfge(m,1)$ in CFG is\ldots & then the VC is equivalent to\ldots \\\hline

$\lassg{n}{x}{e}$ &
$m=\elab(b,c,\fin)$ & $an(n,1)\imp \subst{an(m,1)}{x|}{e|}$ 
\\

$\lifc{n}{e}{b_0}{b_1}$ & $m=\lab(b_0)$ & $an(n,1)\land \leftF{e} \imp an(m,1)$ 
\\

$\lchoice{n}{b_0}{b_1}$ & $m$ is $\lab(b_0)$ or $\lab(b_1)$ & $an(n,1) \imp an(m,1)$ 
\end{tabular}

\medskip

\begin{tabular}{lll}
if $b = \sub(n,d)$ is\ldots & and $(\fin,n)\cfge (\fin,m)$ in CFG is\ldots & then the VC is equivalent to\ldots \\\hline

$\lassg{n}{x}{e}$ &
$m=\elab(b,d,\fin)$ & $an(\fin,n)\imp \subst{an(\fin,m)}{|x}{|e}$ 
\\

$\lifc{n}{e}{b_0}{b_1}$ & $m=\lab(b_0)$ & $an(\fin,n)\land \rightF{e} \imp an(\fin,m)$ 
\\

$\lchoice{n}{b_0}{b_1}$ & $m$ is $\lab(b_0)$ or $\lab(b_1)$ & $an(\fin,n) \imp an(\fin,m)$ 
\end{tabular}

\end{small}
\end{center}
\vspace*{-1ex}
\caption{Selected VCs for sequential product of $\aut(c;\lskipc{\fin})$ and $\aut(d;\lskipc{\fin})$ and full annotation $an$.}\label{tab:VCseq}
\end{figure*}

Let $\Q=an(\fin,1)$.  
We will use $\Q^\eplus$ as the intermediate assertion for rule \rn{Seq} in a proof of
$c;\fdot(d) : \spec{\R^\eplus}{\S^\eplus}$ which can then be used in \rn{SeqProd}
to obtain $c|d : \rspec{\R}{\S}$.   
To obtain proofs of $c:\spec{\R^\eplus}{\Q^\eplus}$ and $\fdot(d):\spec{\Q^\eplus}{\S^\eplus}$,
we use the VCs of Fig.~\ref{tab:VCseq} in an argument similar to the 
proof of Thm.~\ref{thm:floydComplete}.
By induction on $c$ we can show,
for all subcommands $b$ of $c$:
\begin{equation}\label{eq:seqA}
\proves b : \spec{an(\lab(b),1)^\eplus}{an(m,1)^\eplus} 
\end{equation}
where $m=\elab(b,c,\fin)$.
By induction on $d$ we can show,
for all subcommands $b$ of $d$:
\begin{equation}\label{eq:seqB}
\proves \fdot(b) : \spec{an(\fin,\lab(b))^\eplus}{an(\fin,m)^\eplus} 
\end{equation}
where $m=\elab(b,d,\fin)$. 
In proving (\ref{eq:seqB}) using Fig.~\ref{tab:VCseq},
we use that $(an(\fin,n)\land\rightF{e})^\eplus = an(\fin,n)^\eplus\land \fdot(e)$
and $(\subst{an(\fin,n)}{|x}{|e})^\eplus = \subst{(an(\fin,n)^\eplus)}{\Dot{x}}{\fdotsmall(e)}$.  
Instantiating (\ref{eq:seqA}) and (\ref{eq:seqB})
we get
\(\begin{array}[t]{l}
\proves c: \spec{an(1,1)^\eplus}{an(\fin,1)^\eplus} \\
\proves \fdot(d): \spec{an(\fin,1)^\eplus}{an(\fin,\fin)^\eplus} 
\end{array}\)
\\
Thus 
$\proves c:\spec{\R^\eplus}{\Q^\eplus}$ and $\proves \fdot(d):\spec{\Q^\eplus}{\S^\eplus}$,
because $an(1,1)=\R$, $an(\fin,1)=\Q$, and $an(\fin,\fin)=\S$.

\section{A logic of lockstep alignment}\label{sec:lockstep}

\begin{figure*}[t]
\begin{small}
\begin{mathpar}
\inferrule[dSkip]{}{
\skipc \sep \skipc : \rspec{\R}{\R}
}

\inferrule[dAss]{}{
x:=e \sep x':=e' : \rspec{\subst{\R}{x|x'}{e|e'}}{\R}
}

\inferrule*[left=dSeq]{
  c\sep c' : \rspec{\R}{\Q} \\
  d\sep d' : \rspec{\Q}{\S}
}{
  c;d \Sep c';d' : \rspec{\R}{\S}
}

\inferrule*[left=dIf]{
  \R \imp \eqbib{e}{e'} \\
  c\sep c' : \rspec{\R\land \leftF{e}\land\rightF{e'}}{\S} \\
  d\sep d' : \rspec{\R\land \neg\leftF{e}\land\neg\rightF{e'}}{\S} 
}{
  \ifc{e}{c}{d} \Sep \ifc{e'}{c'}{d'} : \rspec{\R}{\S}
}

%
\inferrule*[left=dWh]{
  \Q \imp \eqbib{e}{e'} \\
  c\sep c' : \rspec{\Q\land \leftF{e}\land\rightF{e'}}{\Q} 
}{ 
  \whilec{e}{c} \Sep \whilec{e'}{c'} : \rspec{\Q}{\Q\land \neg\leftF{e}\land\neg\rightF{e'}}
}

\inferrule*[left=rConseq]{
  \P\imp \R \\
  c\sep d : \rspec{\R}{\S} \\
  \S \imp \Q 
}{
  c|d : \rspec{\P}{\Q} \\
}
\end{mathpar}
\end{small}
\vspace*{-3ex}
\caption{Lockstep (diagonal) syntax-directed rules.}\label{fig:lockstep}
\end{figure*}

So far we have that \rn{SeqProd} is complete in the sense of Cook, 
and alignment complete with respect to alignments represented by sequential product.
It is not complete with respect to other classes of alignments.
For example, consider lockstep-control alignments.
As mentioned in Sec.~\ref{sec:intro}, such an alignment enables to prove 
$c0|c0:\rspec{\eqbi{x}{x}}{\eqbi{z}{z}}$ using an annotation
with intermediate relations only $\eqbi{y}{y}\land\eqbi{z}{z}$. 
A proof using \rn{SeqProd} with $c0;\fdot(c0)$ 
requires to assert $z=x! \land x=\Dot{x}$ at the semicolon,
and to use factorial in invariants.
This is far beyond the assertions and judgments associated with the annotation
of the lockstep product.

Fig.~\ref{fig:lockstep} gives rules for relational judgments 
sometimes called ``diagonal''~\cite{Francez83} because they relate same-structured programs.
They are typical of RHLs~\cite{Benton:popl04,Yang:tcs04} and we call them lockstep because they 
embody lockstep-control alignment, 
with side conditions for agreement of tests.
The relational version of \rn{Conseq} is included in Fig.~\ref{fig:lockstep} because
it is needed in order for this collection of rules to be complete for 
lockstep-control alignment,
which we make precise in Theorem~\ref{thm:lckcomplete}.
These rules are not complete in the sense of Cook, for relational 
judgments in general, because they do not apply to differently-structured commands
and do not support reasoning about differing control paths.
For example, the monotonicity property 
$\rspec{x\leq x'}{y\leq y'}$
is satisfied by
$\ifc{x>0}{y:=x+1}{y:=x}$,
but $x\leq x'$ does not imply agreement on the value of $x>0$.

In a lockstep-control product, the CFG edges have the form $(n,n)\cfge(m,m)$.
For this to be sufficient for $\R$-adequacy, 
the code paths reached from initial state-pairs satisfying $\R$ need to be the same.
Thus lockstep control is not adequate for programs with choice 
except in trivial cases like $\lchoice{}{x:=0}{x:=0}$.
Choice is ruled out in theorem.


Say $c$ and $c'$ have \dt{same control}, written $\sameCtl(c,c')$,
if $\labs(c)=\labs(c')$ and for each $n\in\labs(c)$ the programs $\sub(n,c)$ and $\sub(n,c')$ are the same kind:
both are assignments, both are skip, both are if, and so on;  
moreover $n$ has the same control flow successors in $c$ and in $c'$.  
Put differently: $c'$ can be obtained from $c$ by renaming variables and replacing expressions in assignments and branch conditions, but no other changes.
For example, $\lassg{5}{x}{y+z}$ has same control as $\lassg{5}{w}{x-1}$;
so too $c4$ and $c5$ in Fig.~\ref{fig:c4c5}.
Same control implies identical CFGs.  

\begin{figure*}
\begin{small}
\begin{tabular}{lll}
if $b = \sub(n,c)$ and $b' =\sub(n,c')$ are\ldots & and $(n,n)\cfge(m,m)$ in CFG is\ldots & then the VC is equivalent to\ldots \\\hline

$\lassg{n}{x}{e}$ and $\lassg{n}{x'}{e'}$
&
$m=\elab(b,c,\fin)=\elab(b',c',\fin)$ & $an(n,n)\imp \subst{an(m,m)}{x|x'}{e|e'}$ 
\\

$\lskipc{n}$ and $\lskipc{n}$
&
$m=\elab(b,c,\fin)=\elab(b',c',\fin)$ & $an(n,n)\imp an(m,m)$
\\

$\lifc{n}{e}{b_0}{b_1}$ and $\lifc{n}{e'}{b'_0}{b'_1}$ 
& $m=\lab(b_0)=\lab(b_0')$ & $an(n,n)\land \leftF{e}\land\rightF{e'} \imp an(m,m)$ 
\\

$\lifc{n}{e}{b_0}{b_1}$ and $\lifc{n}{e'}{b'_0}{b'_1}$ 
& $m=\lab(b_1)=\lab(b_1')$ & $an(n,n)\land \neg\leftF{e}\land\neg\rightF{e'} \imp an(m,m)$ 

\end{tabular}
\end{small}
\caption{Selected VCs for lockstep-control product of $\aut(c;\lskipc{\fin})$ and $\aut(c';\lskipc{\fin})$
with $\sameCtl(c,c')$, and full annotation $an$.
}\label{tab:VClock}
\end{figure*}



For $c,c'$ with same control, and their lockstep-control product $\Pi$, VCs for a full annotation 
are given in Fig.~\ref{tab:VClock}.
As usual, the VCs for conditional have the test or its negation as antecedent, 
because the VC embodies the program semantics while assuming control is along a particular path; see (\ref{eq:VC}).

Regardless of whether the annotation is full, if 
the cutpoints include all branch conditions, 
and for each point $n$ with branch conditions $e,e'$, respectively,
we have $an(n,n)\imp\eqbib{e}{e'}$,
then $\Pi$ is $\R$-adequate (if $an$ is valid).
This is because by program semantics and definition of lockstep-control, the only stuck non-terminated states are those where
the program is a conditional branch and the conditions disagree (so the successor control points differ).  

As with Prop.~\ref{prop:iamcomplete} and Theorem~\ref{thm:seqcomplete},
an annotation determines a set of what we call associated judgments.
For lockstep automata, the associated judgments are much like those defined
preceding Prop.~\ref{prop:iamcomplete} only doubled, like
$b|b':\rspec{an(\lab(b),\lab(b))}{an(m,m)}$ where $m=\elab(b,c,\fin)$,
together with those obtained by adding if-tests and so forth.
For lack of space we refrain from spelling them out.

\begin{theorem}[alignment completeness for lockstep product]\label{thm:lckcomplete}
Suppose $c$ and $c'$ are choice-free and satisfy $\sameCtl(c,c')$.
Let $\Pi$ be the lockstep-control product of 
$\aut(c;\lskipc{\fin})$ and $\aut(c';\lskipc{\fin})$ and let $an$ be a valid full annotation of $\Pi$ for $\spec{\R}{\S}$.
Assume that for any branch point $n$ with tests $e,e'$
we have $an(n,n)\imp\eqbib{e}{e'}$.
Then $\proves c|c' : \rspec{\R}{\S}$ 
in the logic comprising just the rules of Fig.~\ref{fig:lockstep}.
Moreover this can be proved using only the associated judgments. 
\end{theorem}
Informally, for a relation between two programs with the same control structure, 
one may be able to argue that under precondition $\R$ every pair of executions 
follows the same control path.  To formalize such an argument, 
the annotation at each branch should include that their tests agree.
The theorem says this pattern of reasoning is covered by the rules of Fig.~\ref{fig:lockstep}.

Such reasoning does not apply in the presence of pure nondeterministic choice.
Choice is sometimes used to model externally determined inputs.
An alternative is to model inputs using additional variables,
on which the precondition can assert agreement.

To prove the theorem, we use that $c$ and $c'$ satisfy $\sameCtl(c,c')$. 
We show, by induction on structure of $c$, that
\pagebreak\par
\noindent
for every subprogram $b$ of $c$, with corresponding subprogram $b'$ in $c'$:
\begin{equation}\label{eq:lock}
\proves b \sep b' : \rspec{an(\lab(b),\lab(b))}{an(m,m)} 
\end{equation}
where $m=\elab(b,c,\fin)$. 
Note that $\lab(b)=\lab(b')$ and $m = \elab(b',c',\fin)$ by $\sameCtl(c,c')$.
In the base case, $b$ and $b'$ are both assignments or both skip.
We get that $an(\lab(b),\lab(b))$ implies the weakest precondition for the command to establish $an(m,m)$ by 
the first two rows in Fig.~\ref{tab:VClock}.
So we can use \rn{dSkip} or \rn{dAss}, together with \rn{rConseq}, to get (\ref{eq:lock}).
For the induction step, 
consider the case where 
$b$ is $\lifc{n}{e}{b_0}{b_1}$ and 
$b'$ is $\lifc{n}{e'}{b'_0}{b'_1}$.
Let $m_0=\lab(b_0)=\lab(b'_0)$ (using $\sameCtl$) and 
$m_1=\lab(b_1)=\lab(b'_1)$.
Let $p = \elab(b,c,\fin) = \elab(b',c',\fin)$, 
noting that $p$ is also the end label for the then and else parts.
By induction we have 
$\proves b_0|b'_0 : \rspec{an(m_0,m_0)}{an(p,p)}$
and 
$\proves b_1|b'_1 : \rspec{an(m_1,m_1)}{an(p,p)}$.
By the VCs we have $an(n,n)\land \leftF{e}\land\rightF{e'} \imp an(m_0,m_0)$ 
and $an(n,n)\land \neg\leftF{e}\land\neg\rightF{e'} \imp an(m_1,m_1)$ 
so using \rn{rConseq} we get 
\[ 
\begin{array}[t]{l}
\proves b_0|b'_0 : \rspec{an(n,n)\land \leftF{e}\land\rightF{e'}}{an(p,p)} \\
\proves b_1|b'_1 : \rspec{an(n,n)\land \neg\leftF{e}\land\neg\rightF{e'}}{an(p,p)} 
\end{array}
\]
By assumption of the theorem we have $an(n,n)\imp\eqbib{e}{e'}$,
which is the side condition of rule \rn{dIf},
which yields: 
\begin{small}
\[ \lifc{n}{e}{b_0}{b_1} \Sep \lifc{n}{e'}{b'_0}{b'_1} : \rspec{an(n,n)}{an(p,p)} \]
\end{small}
The arguments for sequence and while are
similar, using rules \rn{dSeq} and \rn{dWh}.  That completes the proof of
(\ref{eq:lock}), as $c$ and $c'$ are choice-free.  Instantiating (\ref{eq:lock})
with $c,c'$ completes the proof.

Theorem~\ref{thm:lckcomplete} pertains to a restricted class of program pairs.
We could relax the $\sameCtl$ condition slightly, to allow an assignment to match skip,
still using lockstep control for the product.
Then the VCs of Fig.~\ref{tab:VClock} would include
a case for $\lassg{n}{x}{e}$ on the left and $\lskipc{n}$ on the right,
with VC  $an(n,n)\imp \subst{an(m,m)}{x|}{e|}$
where $m=\elab(b,c,\fin)=\elab(b',c',\fin)$.
To get an alignment complete logic one would add the axiom
$x:=e \sep \skipc  : \rspec{\subst{\R}{x|}{e|}}{\R}$
(and a similar VC and rule for assignment on the right).

\section{Combining lockstep with sequential}\label{sec:lockstep-seq}

A common practical problem is regression verification: equivalence of two programs that differ in that some subprogram has been replaced by another.  We can describe this as equivalence of $\hat{c}[b]$ and $\hat{c}[b']$,
using the usual notation $\hat{c}[b]$ for a program context $\hat{c}[\,]$ with a designated subprogram $b$.   
Informal reasoning might go by lockstep except for $b$ and $b'$.
In this section we consider a more general situation, relating $\hat{c}[b]$ to $\hat{c'}[b']$ where the contexts $\hat{c}[\,]$ and $\hat{c'}[\,]$ have the same structure, as in Sec.~\ref{sec:lockstep}.
We consider the logic comprised of \rn{SeqProd}, HL, and the rules of Fig.~\ref{fig:lockstep}.
The corresponding form of automata has both lockstep and one-sided sequential steps, where lockstep execution is used for similar control structure and one-sided only for the designated subprograms (which may have arbitrarily different structure).

To be precise, define $\sameExcept(c,c',b,b',beg,end,\fin)$ 
iff there are contexts $\hat{c}[\,]$ and $\hat{c'}[\,]$ such that 
\begin{ditemize}
\item $c = \hat{c}[b]$ and $c' = \hat{c'}[b']$ 
\item $beg=\lab(b)=\lab(b')$ and $\labs(b)\intersect\labs(b') = \{ beg \}$ 
\item $end=\elab(b,c,\fin)=\elab(b',c',\fin)$.  
\item $\sameCtl(\hat{c}[\lskipc{beg}],\hat{c'}[\lskipc{beg}])$ 
\item $\hat{c}[\,]$ and $\hat{c'}[\,]$ are choice free (but $b,b'$ may have choice)
\end{ditemize}

The setup encompasses any pair of programs, because it includes 
the extreme case where $\hat{c}[\,]$ is nothing more than the hole to be filled, i.e.,
$\hat{c}[b] = b$ and $\hat{c'}[b'] = b'$; put differently, $c=b$ and $c'=b'$. 
Here is an example that is only a little beyond what is encompassed by Theorem~\ref{thm:lckcomplete}.
\begin{ditemize}
\item 
$c$ is $\lifc{1}{x>y}{ \lassg{2}{y}{y};\lassg{3}{x}{0} }{\lskipc{4}}$ and 
\item
$c'$ is $\lifc{1}{y\leq x-1}{ \lassg{2}{x}{0} }{\lskipc{4}}$
\item 
$b$ is $\lassg{2}{y}{y};\lassg{3}{x}{0}$ and $b'$ is a single assignment $\lassg{2}{x}{0}$
\end{ditemize}
Here $beg=2$ and $\elab(b,c,\fin)=\fin$.
The example does not satisfy $\sameCtl$ (because $b$ and $b'$ do not match).

To describe the product we use extra control state, for which we assume a set of three tags $\{lck,lo,ro\}$ to designate lockstep, left-only, and right-only steps.
Let $Ctrl$ and $Ctrl'$ be the control sets for the automata of $c;\lskipc{\fin}$
and $c';\lskipc{\fin}$ respectively, assuming as before that $\lab(c)=1=\lab(c')$.
Recall $Ctrl = \labs(c;\lskipc{\fin})$ and  
$Ctrl' = \labs(c';\lskipc{\fin})$.
The control set of the product is $Ctrl \times Ctrl' \times \{lck,lo,ro\}$
and $\LT$ and $\RT$ are the first two projections. 
The initial and final control points are $(1,1,lck)$ and $(\fin,\fin,lck)$.
Define $\biTrans$ as follows, where 
$\trans$ is from  $\aut(c;\lskipc{\fin})$,
$\trans'$ is from  $\aut(c';\lskipc{\fin})$,
and $\biTrans_{lckc}$ is the lockstep-control product based on those.
For all $n,m,s,t,s',t'$:
\begin{small}
\begin{list}{}{}
\item[(i)] 
$((n,n,lck),(s,s')) \biTrans ((m,m,lck),(t,t'))$
\\
if 
$((n,n),(s,s')) \biTrans_{lckc} ((m,m),(t,t'))$
and $n\notin\labs(b)\union\labs(b')$ and $m\neq beg$ 

\item[(ii)]
$((n,n,lck),(s,s')) \biTrans ((beg,beg,lo),(t,t')) $
\\
if 
$((n,n),(s,s')) \biTrans_{lckc} ((beg,beg),(t,t'))$ and
$n\notin\labs(b)\union\labs(b')$ 

\item[(iii)]
$((n,beg,lo),(s,s')) \biTrans ((m,beg,lo),(t,s')) $
\\
if $m\neq end$ and $(n,s) \trans (m,t)$
and $n\in\labs(b)$

\item[(iv)]
$((n,beg,lo),(s,s')) \biTrans ((end,beg,ro),(t,s')) $
\\
if $(n,s) \trans (end,t)$
and $n\in\labs(b)$

\item[(v)]
$((end,n',ro),(s,s')) \biTrans ((end,m',ro),(s,t')) $
\\
if $m'\neq end$ and $(n',s') \trans' (m',t')$
and $n'\in\labs(b')$

\item[(vi)] 
$((end,n',ro),(s,s')) \biTrans ((end,end,lck),(s,t')) $ \\
if $(n',s') \trans' (end,t')$
and $n'\in\labs(b)$
\end{list}
\end{small}
Rule (ii) enters left-only mode, (iii) continues, (iv) switches to right-only, 
(v) continues, and (vi) resumes lockstep.  
Like the lockstep-control product, this gets stuck at branch points outside the designated subprograms $b,b'$, if tests don't agree.

\begin{theorem}\label{thm:lockstep-seq}
Suppose $\sameExcept(c,c',b,b',beg,end,\fin)$  and $\Pi$ is the product defined above.
Suppose $an$ is a valid full annotation of $\Pi$ for $\spec{\R}{\S}$.
Suppose for all branch points $n\in\labs(c)\setminus(\labs(b)\union\labs(b'))$,
with branch conditions $e,e'$, we have $an(n,n,lck)\imp\eqbib{e}{e'}$.
Then $\proves c|c' : \rspec{\R}{\S}$ 
in the logic comprised of HL, the rules of Fig.~\ref{fig:lockstep}, and \rn{SeqProd},
using only the associated judgments. 
\end{theorem}

As an exercise the reader may like to modify the product in this section 
to handle the special case where 
$b$ is $\choice{b0}{b1}$ and $b'$ is $\choice{b0'}{b1'}$
by nondeterministically choosing between
four sequential executions
($b0'$ after $b0$, or $b1'$ after $b0$, etc).
Then recover alignment completeness by adding 
a relational proof rule that relates a choice to a choice, with four premises.

\section{Conditionally aligned loops}\label{sec:dissonant}


We want to prove 
$c2$ majorizes $c3$, that is, the judgment 
$c2\sep c3 : \rspec{x = x'\land x > 3}{z>z'}$, by aligning the iterations in which $y$ gets updated, maintaining invariant $y=y' \land z>z'$ and allowing the no-op iterations to happen independently. 
To do so we use this rule~\cite{Beringer11}:

\[
\inferrule[caWhile]{
  c\sep c' : \rspec{\Q\land \leftF{e}\land\rightF{e'}\land\neg\Lrel\land\neg\R}{\Q} \\
  c\sep\skipc : \rspec{\Q\land\Lrel\land \leftF{e}}{\Q} \\
  \skipc\sep c' : \rspec{\Q\land\R\land \rightF{e'}}{\Q} \\
  \Q \imp \eqbib{e}{e'} \lor (\Lrel\land\leftF{e}) \lor (\R\land\rightF{e'})  
}{ 
  \whilec{e}{c} \Sep \whilec{e'}{c'} : \rspec{\Q}{\Q\land \neg\leftF{e}\land\neg\rightF{e'}}
}
\]
There are three premises, which strengthen the invariant $\Q$ in three different ways.
The first premise relates both loop bodies, like the lockstep loop rule in Fig.~\ref{fig:lockstep}.  The second and third premises each relate a loop body to skip,
under preconditions strengthened by relations $\Lrel$ or  $\R$.
The side condition ensures these are adequate to cover all cases.
Later we consider the example of $c2,c3$ using $\leftF{w\mod 2\neq 0}$ for $\Lrel$
and $\rightF{w'\mod 3\neq 0}$ for $\R$.
For \rn{caWhile} to be useful, the proof system we consider has 
the lockstep rules of Fig.~\ref{fig:lockstep} together with one-sided rules of Fig.~\ref{fig:one-side}.

In order to focus on this situation, we consider relating 
same-control programs $c,c'$ with a distinguished label $beg$ 
such that $\sub(beg,c)$ is a loop---and so is $\sub(beg,c')$, since we assume 
$\sameCtl(c,c')$.
Let $Ctrl,Ctrl'$ and $\trans,\trans'$ be the control sets and transition relations for their automata
(noting that $Ctrl'=Ctrl$).
As in Sec.~\ref{sec:lockstep-seq} we define a product 
with control $Ctrl\times Ctrl'\times Tag$ where $Tag = \{lck,lo,ro\}$.
The transition relation $\biTrans$ is defined with respect to 
given relations $\Lrel$ and $\R$.
Unlike a ctrl-conditioned automaton, some transitions are conditioned on the stores.
If $n$ is the label of the distinguished loop's body, 
there are three transitions from the top of the loop into its body:
$(beg,beg,lck)\cfge (n,n,lck)$ if neither $\Lrel$ nor $\R$ holds,
$(beg,beg,lck)\cfge (n,beg,lo)$ if $\Lrel$ holds,
and 
$(beg,beg,lck)\cfge (beg,n,ro)$ if $\R$ holds.
Fig.~\ref{fig:aut3} depicts an example.
\iflongver
(See appendix for details.)
\fi

\begin{theorem}\label{thm:lockstep-While3}
Consider $c,c',beg,\Lrel,\R$ such that 
$\sameCtl(c,c')$, $c$ and $c'$ are choice-free, 
$\sub(beg,c)$ is a loop, 
and $\Lrel,\R$ are store relations.
Let $\Pi$ be the product described above for $c;\lskipc{\fin}$, $c';\lskipc{\fin}$.
Suppose $an$ is a valid full annotation of $\Pi$ for $\spec{\P}{\Q}$.
Assume 
\begin{list}{}{}
\item[(a)] for all branch points $n\in\labs(c)\setminus\{beg\}$  
with branch conditions $e,e'$, we have $an(n,n,lck)\imp\eqbib{e}{e'}$; and
\item[(b)] $an(beg,beg,lck)\imp \eqbib{e}{e'} \lor (\Lrel\land\leftF{e'}) \lor (\R\land\rightF{e})$
where $e,e'$ are the tests of the loops at $beg$.
\end{list}
Then $\proves c\sep c' : \rspec{\P}{\Q}$ 
in the logic comprised of 
the rules of Fig.~\ref{fig:lockstep}, Fig.~\ref{fig:one-side},
and \rn{caWhile}, 
using only the associated judgments. 
\end{theorem}
Assumption (a) is about annotations of aligned branch points,
even inside the distinguished loop at $beg$, but only 
in the part of the product CFG that executes in lockstep.
For $n$ a branch point inside that loop, the control points $(n,n,lo)$ and $(n,n,ro)$ 
are not reachable, by construction of $\Pi$.  
Assumption (b) is like the side condition of rule \rn{caWhile} and ensures adequacy.

\begin{figure}[t]
\begin{small}
\begin{mathpar}

\inferrule[AssSkip]{}{
x:=e \sep \skipc  : \rspec{\subst{\R}{x|}{e|}}{\R}
}

\inferrule[SkipSkip]{}{
\skipc \sep \skipc  : \rspec{\R}{\R}
}

\inferrule*[left=SeqSkip]{
  c \sep\skipc : \rspec{\R}{\Q} \\
  d \sep \skipc : \rspec{\Q}{\S} 
}{
  c;d \sep \skipc : \rspec{\R}{\S}
}

\inferrule*[left=IfSkip]{
  c\sep \skipc : \rspec{\R\land \leftF{e}}{\S} \\
  d\sep \skipc : \rspec{\R\land \leftF{\neg e}}{\S} 
}{
  \ifc{e}{c}{d} \Sep \skipc : \rspec{\R}{\S}
}

\inferrule*[left=WhSkip]{
  c\sep \skipc : \rspec{\Q\land \leftF{e}}{\Q} 
}{ 
  \whilec{e}{c} \Sep \skipc : \rspec{\Q}{\Q\land \leftF{\neg e}}
}
\end{mathpar}
\end{small}
\vspace*{-3ex}
\caption{One-side rules (symmetric right-side rules omitted).}
\label{fig:one-side}
\end{figure}

Instead of showing how $c2$ majorizes $c3$, we consider variations $c4,c5$ in Fig.~\ref{fig:c4c5}.
This lets us avoid complications in the invariant needed to handle the first few iterations 
of $c2,c3$ (where $z>z'$ does not hold under precondition $x=x'$).
Whereas the originals maintain the invariants 
$x! = z * y!$ (for $c2$) and $2^x = z * 2^y$ (for $c3$) and $y\geq 0$ (for both), 
the variations maintain $x!*4! = z * y!$ and $2^x*2^4 = z * 2^y$ and $y\geq 4$,
owing to initializations of $z=4!=24$ and $z=2^4=16$.
We shall prove $\rspec{x = x'\land x > 3}{z>z'}$
for $c4;\lskipc{0}$ and $c5;\lskipc{0}$.
We use the product construction of this section,
for alignment conditions mentioned earlier:
$\Lrel$ is $\leftF{ w\mod 2\neq 0 }$
and $\R$ is $\rightF{  w'\mod 3\neq 0 }$.
So the transition on edge
$(4,4,lck)\cfge(5,4,lo)$ 
is guarded by $\leftF{ w\mod 2\neq 0}$,
the transition
$(4,4,lck)\cfge(4,5,ro)$ 
is guarded by $\rightF{w'\mod 3\neq 0}$,
and $(4,4,lck)\cfge(5,5,lck)$ is guarded by $\neg\Lrel\land\neg\R$.
As loop invariant we choose 
\[ \S: \qquad y=y' \land y>3 \land z>z'>0 \]
Define the annotation $an$ as follows.
\begin{small}
\[  
\begin{array}{ll}
(n,m,tag) & an(n,m,tag) \\\hline
(1,1,lck) & x=x'\land x>3 \qquad\mbox{precondition}\\
(2,2,lck)  & y=y' \land y>3 \\
(3,3,lck)  & y=y' \land y>3 \land z>z'>0 \\
(4,4,lck)  & \S \\
(5,5,lck)  & \S \land y>4 \land \neg\Lrel\land\neg\R\\
(6,6,lck)  & \S \land y>4 \land w\mod 2 = 0 = w'\mod 3 \\ 
(7,7,lck)  & \S \land y>4 \land w\mod 2 = 0 = w'\mod 3 \\ 
(8,8,lck)  & \S \land y>4 \land w\mod 2 \neq 0 \neq w'\mod 3 \\ 
  (9,9,lck)  & \S \\ 
(5,4,lo)  & \S \land \Lrel  \\
(8,4,lo)  & \S   \\
(9,4,lo)  & \S   \\
(4,5,ro)  & \S \land \R  \\
(4,8,ro)  & \S    \\
(4,9,ro)  & \S   \\
(0,0,lck) & z>z' \qquad\mbox{postcondition}
\end{array}
\]
\end{small}

Let $an(m,n,tag)=false$ for all others.
The automaton never reaches $(6,8,lck)$ or $(8,6,lck)$,
owing to the lockstep conditions.
It never reaches $(6,4,lo)$ or $(7,4,lo)$ because $\Lrel$ contradicts the if-test;
likewise $(4,6,ro)$, $(4,7,ro)$, and $\R$.
The annotation is valid.



\section{Discussion}\label{sec:discuss}

We introduced a notion of completeness relative to a designated class of alignments, and 
showed alignment completeness for four illustrative sets of RHL rules.
In passing, we defined and proved Floyd completeness for HL.
For simplicity we used the basic semantics of specs. 
Non-stuck semantics is preferable for richer languages 
and the requisite adjustments to HL and RHL rules are straightforward and well known 
(e.g., the precondition for assignment includes a definedness condition).  
We highlighted what adequacy means for non-stuck semantics, but refrained from spelling out alignment completeness results for it.

In this section we point out related work and directions for further development.
For more extensive reviews of related work on RHLs, some starting points
are~\cite{BeckertU18short,BartheCK16,MaillardHRM20} and~\cite{NaumannISOLA20}.  
Naumann~\cite{NaumannISOLA20} proposes the idea of alignment completeness but only in vague terms.

Product automata appear in many places (e.g.,\cite{Francez83,BartheCK13,ChurchillP0A19})
and are similar to control flow automata~\cite{HeizmannHP13,LangeEtalFMCAD15}.
Francez~\cite{Francez83} observes that sequential product is complete relative to HL,
but does not work that out in detail.
Its completeness is featured in Barthe et al.~\cite{BartheDArgenioRezk}, for the special case of relating
a program to itself; it is clear that it holds generally as noted in~\cite{BartheCK-FM11}.
Beringer~\cite{Beringer11} proves semantic completeness of sequential product
and leverages it to derive rules including two conditionally aligned loop rules. 
The variation \rn{caWhile} is featured in Banerjee et al.~\cite{BanerjeeNN16short,NikoueiBN19a};
the latter uses a form of dovetail product for non-stuck semantics.  
A RHL for deterministic programs is proved complete based on sequential product by Barthe et al.~\cite{BartheGHS17}.
Sousa and Dillig~\cite{SousaD16} give a rule like \rn{SeqProd} for $k$-products;
their Theorem 2 is completeness relative to completeness of an underlying HL.
Wang et al.~\cite{WangDilligLahiriCook} prove a similar result specialized to program equivalence.
The basis of these results is that the product under consideration is adequate, to use our term.
Francez gives an adequacy result of this sort, for eager-lockstep,
and Eilers et al.~\cite{EilersMH20} do the same for a more general form of $k$-product
that uses eager-lockstep for loops.
Aguirre et al.~\cite{AguirreBGGS19} prove completeness of a
RHL for higher order functional programs, via embedding in a unary logic.

By contrast with these Cook-style completeness results, alignment completeness is with respect to 
a designated class of alignments. Our results are for classes of alignments defined in terms of
a limited class of product automata without auxiliary store,
although ghost variables can be used in the relations $\Lrel$ and $\R$ of rule \rn{caWhile}.  
The most general notion of alignment would be a function from pairs of traces to alignments thereof~\cite{KovacsSF13}.
Even restricted to computable functions, this is far beyond the scope of known RHL rules,
even using mixed-structure rules like this one adapted from~\cite{Francez83}.
\[
\inferrule{  
\whilec{e\land e_0 }{b} \sep c : \rspec{\P}{\Q} \\
\whilec{e}{b} \sep d : \rspec{\Q}{\R} \\
\Q\land\leftF{\neg e}\imp\R
}{
\whilec{e}{b} \Sep c;d : \rspec{\P}{\R}
}
\]
Several practical works use more sophisticated product constructions~\cite{ChurchillP0A19,ShemerGSV19}, 
including some that use auxiliary store~\cite{GirkaEtalPPDP17,ClochardMP20}.
Clochard et al.~\cite{ClochardMP20} highlight the efficacy of (unary) deductive verification applied to product programs, as well as going beyond $\forall\forall$ (as do~\cite{BartheCK13}).
Bringing more sophisticated alignments within the purview of RHL could have the familiar benefits of HL,
like bringing the principles of conjunctive and disjunctive decomposition together with the IAM.   
However, merely collecting a large number of mixed-structure and data-conditioned rules would be inelegant
and likely fall short of covering all useful alignments.     
An alternative is to leverage HL in the way \rn{SeqProd} does, but for a wider range of product encodings.  
This might be achieved using a subsidiary judgment that 
connects two commands with a third that is an adequate product, 
as explored in~\cite{BartheCK16,NikoueiBN19a},
and taking into account the encoding of two stores by one for different data models~\cite{BartheDArgenioRezk,Naumann06esorics,BeringerHofmann07}.

Another subsidiary judgment that broadens the applicability of basic RHL rules 
is correctness-preserving rewriting,
which is common in verification tools but is seldom made explicit in HLs.
The RHL of~\cite{BanerjeeNN16short} includes an unconditional rewriting relation $\uequiv$
and rule to infer 
$c|c':\rspec{\R}{\S}$ from $d|d':\rspec{\R}{\S}$ if $c\uequiv d$ and $c'\uequiv d'$.
Using simple equivalences like $\ifc{E}{C}{\skipc} ; \whilec{E}{C} \uequiv \whilec{E}{C}$,
together with a version of \rn{caWhile}, they prove a loop tiling transformation which had 
been used to argue for working directly with automata~\cite{BartheCK13}. 
Unrolling  examples $c2$ and $c3$ a few times would address the problem we dodged by using $c4$ and $c5$.
Equivalences valid in Kleene algebra with tests~\cite{Kozen00} 
are a natural candidate to connect with some class of product automata.  

Alignment completeness characterizes sets of rules in terms of classes of alignments.
If different classes could be defined using combinators for product automata,
one might obtain a more uniform and comprehensive theory leading to more systematic 
development of relational verification tools.



\subsection*{Acknowledgments}

Thanks to Anindya Banerjee 
and anonymous reviewers for helpful suggestions.
The authors were partially supported by NSF award CNS 1718713
and the second author was partially supported by 
ONR N00014-17-1-2787.


\bibliographystyle{IEEEtran}
\bibliography{new.bib,naumann.bib}

\begin{thebibliography}{10}
\providecommand{\url}[1]{#1}
\csname url@samestyle\endcsname
\providecommand{\newblock}{\relax}
\providecommand{\bibinfo}[2]{#2}
\providecommand{\BIBentrySTDinterwordspacing}{\spaceskip=0pt\relax}
\providecommand{\BIBentryALTinterwordstretchfactor}{4}
\providecommand{\BIBentryALTinterwordspacing}{\spaceskip=\fontdimen2\font plus
\BIBentryALTinterwordstretchfactor\fontdimen3\font minus
  \fontdimen4\font\relax}
\providecommand{\BIBforeignlanguage}[2]{{%
\expandafter\ifx\csname l@#1\endcsname\relax
\typeout{** WARNING: IEEEtran.bst: No hyphenation pattern has been}%
\typeout{** loaded for the language `#1'. Using the pattern for}%
\typeout{** the default language instead.}%
\else
\language=\csname l@#1\endcsname
\fi
#2}}
\providecommand{\BIBdecl}{\relax}
\BIBdecl

\bibitem{Benton:popl04}
N.~Benton, ``Simple relational correctness proofs for static analyses and
  program transformations,'' in \emph{{POPL}}, 2004.

\bibitem{TerauchiAiken}
T.~Terauchi and A.~Aiken, ``Secure information flow as a safety problem,'' in
  \emph{Static Analysis Symposium ({SAS})}, 2005.

\bibitem{Cook78}
S.~A. Cook, ``Soundness and completeness of an axiom system for program
  verification,'' \emph{{SIAM} J. Comput.}, vol.~7, no.~1, 1974.

\bibitem{AptOld3}
K.~R. Apt, F.~S. de~Boer, and E.-R. Olderog, \emph{Verification of Sequential
  and Concurrent Programs}, 3rd~ed.\hskip 1em plus 0.5em minus 0.4em\relax
  Springer, 2009.

\bibitem{NipkowCSL02}
T.~Nipkow, ``Hoare logics for recursive procedures and unbounded
  nondeterminism,'' in \emph{Computer Science Logic}, 2002.

\bibitem{GodlinS09}
B.~Godlin and O.~Strichman, ``Regression verification,'' in \emph{46th ACM
  Design Automation Conference}, 2009.

\bibitem{Floyd67}
R.~Floyd, ``Assigning meaning to programs,'' in \emph{Symposium on Applied
  Mathematics 19, Mathematical Aspects of Computer Science}, 1967.

\bibitem{TuringChecking49}
A.~Turing, ``On checking a large routine,'' in \emph{Report of a Conference on
  High Speed Automatic Calculating Machines}, 1949.

\bibitem{AptO19}
K.~R. Apt and E.~Olderog, ``Fifty years of {Hoare's} logic,'' \emph{Formal Asp.
  Comput.}, vol.~31, no.~6, 2019.

\bibitem{OwickiGries}
S.~Owicki and D.~Gries, ``An axiomatic proof technique for parallel programs,''
  \emph{Acta Inf.}, vol.~6, 1976.

\bibitem{Beringer11}
L.~Beringer, ``Relational decomposition,'' in \emph{Interactive Theorem Proving
  {(ITP)}}, 2011.

\bibitem{Francez83}
N.~Francez, ``Product properties and their direct verification,'' \emph{Acta
  Informatica}, vol.~20, 1983.

\bibitem{Yang:tcs04}
H.~Yang, ``Relational separation logic,'' \emph{Theo. Comp. Sci.}, vol. 375,
  2007.

\bibitem{BeckertU18short}
B.~Beckert and M.~Ulbrich, ``Trends in relational program verification,'' in
  \emph{Principled Software Development}, 2018.

\bibitem{BartheCK16}
G.~Barthe, J.~M. Crespo, and C.~Kunz, ``Product programs and relational program
  logics,'' \emph{J. Logical and Algebraic Methods in Programming}, vol.~85,
  no.~5, 2016.

\bibitem{MaillardHRM20}
K.~Maillard, C.~Hrit\c{c}u, E.~Rivas, and A.~V. Muylder, ``The next 700
  relational program logics,'' \emph{Proc. {ACM} Program. Lang.}, vol.~4, no.
  {POPL}, 2020.

\bibitem{NaumannISOLA20}
D.~A. Naumann, ``Thirty-seven years of relational {Hoare} logic: remarks on its
  principles and history,'' in \emph{{ISOLA}}, 2020, extended version at
  \url{https://arxiv.org/abs/2007.06421}.

\bibitem{BartheCK13}
G.~Barthe, J.~M. Crespo, and C.~Kunz, ``Beyond 2-safety: Asymmetric product
  programs for relational program verification,'' in \emph{{LFCS}}, 2013.

\bibitem{ChurchillP0A19}
B.~R. Churchill, O.~Padon, R.~Sharma, and A.~Aiken, ``Semantic program
  alignment for equivalence checking,'' in \emph{{PLDI}}, 2019.

\bibitem{HeizmannHP13}
M.~Heizmann, J.~Hoenicke, and A.~Podelski, ``Software model checking for people
  who love automata,'' in \emph{{CAV}}, 2013.

\bibitem{LangeEtalFMCAD15}
T.~Lange, M.~R. Neuh{\"{a}}u{\ss}er, and T.~Noll, ``{IC3} software model
  checking on control flow automata,'' in \emph{{FMCAD}}, 2015.

\bibitem{BartheDArgenioRezk}
G.~Barthe, P.~R. D'Argenio, and T.~Rezk, ``Secure information flow by
  self-composition,'' in \emph{{IEEE CSFW}}, 2004, see extended
  version~\cite{BartheDR11}.

\bibitem{BartheCK-FM11}
G.~Barthe, J.~M. Crespo, and C.~Kunz, ``Relational verification using product
  programs,'' in \emph{Formal Methods}, 2011.

\bibitem{BanerjeeNN16short}
A.~Banerjee, D.~A. Naumann, and M.~Nikouei, ``Relational logic with framing and
  hypotheses,'' in \emph{{FSTTCS}}, 2016, technical report at
  \url{https://arxiv.org/abs/1611.08992}.

\bibitem{NikoueiBN19a}
A.~Banerjee, R.~Nagasamudram, M.~Nikouei, and D.~A. Naumann, ``A relational
  program logic with data abstraction and dynamic framing,'' 2019, available at
  \url{https://arxiv.org/abs/1910.14560}.

\bibitem{BartheGHS17}
G.~Barthe, B.~Gr{\'{e}}goire, J.~Hsu, and P.~Strub, ``Coupling proofs are
  probabilistic product programs,'' in \emph{{POPL}}, 2017.

\bibitem{SousaD16}
M.~Sousa and I.~Dillig, ``Cartesian {H}oare {L}ogic for verifying k-safety
  properties,'' in \emph{{PLDI}}, 2016.

\bibitem{WangDilligLahiriCook}
Y.~Wang, I.~Dillig, S.~K. Lahiri, and W.~R. Cook, ``Verifying equivalence of
  database-driven applications,'' \emph{Proc. {ACM} Program. Lang.}, vol.~2,
  no. {POPL}, 2018.

\bibitem{EilersMH20}
M.~Eilers, P.~M{\"{u}}ller, and S.~Hitz, ``Modular product programs,''
  \emph{{ACM} Trans. Program. Lang. Syst.}, vol.~42, no.~1, 2020.

\bibitem{AguirreBGGS19}
A.~Aguirre, G.~Barthe, M.~Gaboardi, D.~Garg, and P.~Strub, ``A relational logic
  for higher-order programs,'' \emph{J. Funct. Program.}, vol.~29, 2019.

\bibitem{KovacsSF13}
M.~Kov{\'a}cs, H.~Seidl, and B.~Finkbeiner, ``Relational abstract
  interpretation for the verification of 2-hypersafety properties,'' in
  \emph{{CCS}}, 2013.

\bibitem{ShemerGSV19}
R.~Shemer, A.~Gurfinkel, S.~Shoham, and Y.~Vizel, ``Property directed self
  composition,'' in \emph{{CAV}}, 2019.

\bibitem{GirkaEtalPPDP17}
T.~Girka, D.~Mentr\'{e}, and Y.~R\'{e}gis-Gianas, ``Verifiable semantic
  difference languages,'' in \emph{{PPDP}}, 2017.

\bibitem{ClochardMP20}
M.~Clochard, C.~March{\'{e}}, and A.~Paskevich, ``Deductive verification with
  ghost monitors,'' \emph{Proc. {ACM} Program. Lang.}, vol.~4, no. {POPL},
  2020.

\bibitem{Naumann06esorics}
D.~A. Naumann, ``From coupling relations to mated invariants for secure
  information flow,'' in \emph{{ESORICS}}, 2006.

\bibitem{BeringerHofmann07}
L.~Beringer and M.~Hofmann, ``Secure information flow and program logics,'' in
  \emph{{IEEE CSF}}, 2007.

\bibitem{Kozen00}
D.~Kozen, ``On {Hoare} logic and {Kleene} algebra with tests,'' \emph{{ACM}
  Trans. Comput. Log.}, vol.~1, no.~1, 2000.

\bibitem{BartheDR11}
G.~Barthe, P.~R. D'Argenio, and T.~Rezk, ``Secure information flow by
  self-composition,'' \emph{Math. Struct. Comput. Sci.}, vol.~21, no.~6, 2011.

\end{thebibliography}

\iflongver
\else
\end{document}
\fi

\appendix



\subsection*{Definitions}

Definition of $\destutter(\tau)$ for trace $\tau$:
\[\begin{array}{l}
\destutter(\tau) = \tau  \mbox{ if $len(\tau)=1$} \\
\destutter(\gamma_0::\gamma_1::\tau) = \destutter(\gamma_1::\tau) \mbox{ if $\gamma_0 = \gamma_1$} \\
\destutter(\gamma_0 :: \gamma_1 :: \tau) = \gamma_0::\destutter(\gamma_1::\tau) \mbox{ otherwise}
\end{array}
\]

Definition of $\lab(c)$:
\begin{footnotesize}
\[
\begin{array}{lcl}
\lab(\lskipc{n})            & = & n \\
\lab(\lassg{n}{x}{e})       & = & n \\
\lab(c;d)                 & = & \lab(c) \\
\lab(\lifc{n}{e}{c}{d})   & = & n \\
\lab(\lwhilec{n}{e}{c})    & = & n 
\\
\lab(\lchoice{n}{c}{d}) & = & n
\end{array}
\]
\end{footnotesize}

Definition of $\labs(c)$:
\begin{footnotesize}
\[
\begin{array}{lcl}
\labs(\lskipc{n})           & = & \{ n \}\\
\labs(\lassg{n}{x}{e})      & = &\{ n\} \\
\labs(c;d)                & = & \labs(c)\union \labs(d) \\
\labs(\lifc{n}{e}{c}{d})  & = & \{n\}\union \labs(c) \union \labs(d) \\
lab(\lwhilec{n}{e}{c})    & = & \{n\}\union \labs(c) 
\\
\labs(\lchoice{n}{c}{d})  & = & \{n\}\union \labs(c) \union \labs(d) \\
\end{array}
\]
\end{footnotesize}

\begin{figure}[t]
\begin{footnotesize}
\(
\begin{array}{lcl}
\ok(\lskipc{n})            & = & n\geq 0 \\
\ok(\lassg{n}{x}{e})       & = & n\geq 0 \\
\ok(c;d)                 & = & \labs(c)\intersect \labs(d)=\emptyset\land \ok(c)\land \ok(d) \\
\ok(\lwhilec{n}{e}{c})    & = & n\geq 0 \land n \notin \labs(c) \land \ok(c)
\\
\ok(\lifc{n}{e}{c}{d})   & = & 
             n\geq 0 
     \land n\notin \labs(c;d) \land \ok(c;d) 
\\
\ok(\lchoice{n}{c}{d})   & = & 
             n\geq 0 
     \land n\notin \labs(c;d) \land \ok(c;d) 
\end{array}
\)
\end{footnotesize}
\vspace*{-2ex}
\caption{$\ok(c)$ -- command with unique, non-negative labels.}\label{fig:okcmd}
\end{figure}

\begin{figure}[t]
\begin{footnotesize}
\(
\begin{array}{lcl}
\sub(m,\lskipc{m})        & = & \lskipc{m} \\
\sub(m,\lassg{m}{x}{e})   & = & \lassg{m}{x}{e} \\
\sub(m,c;d)               & = & \sub(m,c)\mbox{ , if $m\in \labs(c)$}  \\
                          & = & \sub(m,d)\mbox{ , otherwise} \\
\sub(m, \lwhilec{n}{e}{c}) & = &  \lwhilec{n}{e}{c} \mbox{ , if $n=m$}  \\
                           & = &  \sub(m,c) \mbox{ , otherwise}  \\
\sub(m,\lifc{n}{e}{c}{d}) & = & \lifc{n}{e}{c}{d} \mbox{ , if $n=m$} \\
                 & = & \sub(m,c) \mbox{ , if $m\in \labs(c)$ }\\
                 & = & \sub(m,d) \mbox{ , otherwise } 
\\
\sub(m,\lchoice{n}{c}{d}) & = & \lchoice{n}{c}{d} \mbox{ , if $n=m$} \\
                 & = & \sub(m,c) \mbox{ , if $m\in \labs(c)$ }\\
                 & = & \sub(m,d) \mbox{ , otherwise } 
\end{array}
\)
\end{footnotesize}
\vspace{-2ex}
\caption{$\sub(m,c)$ sub-command of $c$ at $m$,
assuming 
$\ok(c)$,
$ m\in \labs(c)$.}\label{fig:def-sub}
\end{figure}

Fig.~\ref{fig:okcmd} defines the $\ok$ condition.

Fig.~\ref{fig:def-sub} defines $\sub(m,c)$.

As an example how labels are manipulated in the transition semantics,
here is an $\ok$ command and its transitions.
We omit stores but assume $x=1$ initially.
\begin{footnotesize}
\[ \begin{array}{r}
\lwhilec{1}{x>0}{\lassg{2}{x}{x-1}} ; \lskipc{3}  \hfill \\
\ctrans
\lassg{2}{x}{x-1}; \lwhilec{1}{x>0}{\lassg{2}{x}{x-1}} ; \lskipc{3} \\
\ctrans
\lskipc{-2}; \lwhilec{1}{x>0}{\lassg{2}{x}{x-1}} ; \lskipc{3} \\
\ctrans
\lwhilec{1}{x>0}{\lassg{2}{x}{x-1}} ; \lskipc{3} \\
\ctrans
\lskipc{-1};\lskipc{3} \\
\ctrans
\lskipc{3} \\
\end{array}
\]
\end{footnotesize}

\begin{figure}[t]
\begin{small}
\begin{mathpar}

\inferrule[Skip']{P\imp Q}{ \skipc:\spec{P}{Q} }

\inferrule[Ass']{ P\imp \subst{Q}{x}{e} }{ x:=e : \spec{P}{Q} }

\inferrule[If']
{ P\land e \imp P_0 \\ P\land \neg e \imp P_1 \\
c:\spec{P_0}{Q} \\ d:\spec{P_1}{Q}
}{
\ifc{e}{c}{d} : \spec{P}{Q} 
}

\inferrule[Wh']
{ P\land e \imp P_0 \\
  c:\spec{P_0}{P} \\
  P\land \neg e \imp Q 
}{
\whilec{e}{c} : \spec{P}{Q} 
}
\end{mathpar}
\end{small}
\vspace*{-2ex}
\caption{\rn{Conseq}-free HL: these rules plus \rn{Seq} and \rn{Choice}.}\label{fig:minHL}
\end{figure}

Apropos the Floyd completeness Thm.~\ref{thm:floydComplete},
one can dispense with all forms of associated judgment except (\ref{eq:anno})
by reformulating some of the HL rules to more directly match the judgments of a valid annotation so rule \rn{Conseq} becomes unnecessary, with requisite implications added as side conditions of the syntax-directed rules.
This is spelled out in Fig.~\ref{fig:minHL}, and is akin to formal 
systems for proof outlines.  The alternatives have a parallel in type systems,
where subtyping is  handled either with a separate rule of subsumption or 
with subtype constraints in syntax-directed rules that are 
considered algorithmic.  
We choose to formulate our results to fit 
usual presentations of HL and RHL in the literature, at the cost of 
some clutter in the definition of associated judgment for each of the theorems.


\subsection*{Proof of Prop.~\ref{prop:iamsound}}

The second statement, $A\models \spec{P}{Q}$,
follows directly from the first, using that $\fin$ is in $K$.
The proof of the first statement is by induction on the number of $K$-states reached in $\tau$.
At least one is reached, because $\tau$ is non-empty and initial, and $K$ contains $\init$.
If exactly one is reached, we have $\tau_0\models P$ since $an(\init)=P$,
so the base case is done.
Otherwise, suppose a $K$-state is reached at position $j$ and the preceding $K$-state is at position $i$ (so $i<j$).
By induction we have $\tau_i\models an(\ctrl(\tau_i))$.
Let $\tau[i:j]$ be the trace segment from $\tau_i$ to $\tau_j$, inclusive.
Now $\cpath(\tau[i:j])$ is in $\segs(A,K)$ so we can apply the 
verification condition (\ref{eq:VC}) to conclude $\tau_j\models an(\ctrl(\tau_j))$.

\subsection*{Proof of Prop.~\ref{prop:iamcomplete}}

For each $n\in K$, let $an(n)$ be the set of $s$ 
such that there is a finite initial trace $\tau$ of $A$ with $\tau_0\models P$ and $\last{\tau}=(n,s)$.
So $an(n)$ is the strongest invariant at $n$ for traces from $P$.
To prove (\ref{eq:VC}), consider any $vs\in\segs(A,K)$ and trace $\tau$ along $vs$.
Suppose $\tau_0\models an(vs_0)$, to show $\last{\tau}\in an(\last{vs})$.
By definition of $an(vs_0)$, $\tau_0$ is reachable from an initial $P$-state, say by trace $\upsilon$
with $\last{\upsilon}=\tau_0$.
So $\last{\tau}$ is reachable by the trace $\upsilon$ followed by $\tau$, hence
$\last{\tau}\in an(\last{vs})$ by definition of $an$.

\subsection*{Proof of Thm.~\ref{thm:floydComplete}: other cases}

\begin{ditemize}
\item If $b$ is $\lwhilec{n}{e}{d}$,
we have $\elab(d,c,\fin) = n$ so by induction we have
\ghostbox{$ \proves d : \spec{ an(\lab(d)) }{ an(n) } $}.
Lemma~\ref{lem:VCprog} gives $an(n)\land e \imp an(\lab(d))$ 
so by \rn{Conseq} we get 
\ghostbox{$\proves d : \spec{ an(n)\land e }{ an(n) }$}.
Rule \rn{Wh} yields 
\ghostbox{$\proves b : \spec{ an(n) }{ an(n)\land \neg e }$}.
Lemma~\ref{lem:VCprog} gives $an(n)\land \neg e \imp an(\elab(b,c,\fin))$ so by Conseq
we get 
\ghostbox{$\proves b : \spec{ an(n) }{ an(\elab(b,c,\fin) }$}.

\item If $b$ is $\lchoice{n}{d_0}{d_1}$,    by induction we have 
\ghostbox{$\proves d_0 : \spec{ an(\lab(d_0) }{ an(\elab(d_0,c,\fin)) }$}
and 
\ghostbox{$\proves d_1 : \spec{ an(\lab(d_1) }{ an(\elab(d_1,c,\fin)) }$}.
Lemma~\ref{lem:VCprog} gives
$an(n) \imp an(\lab(d_0))$ and $an(n) \imp an(\lab(d_1))$  
so by \rn{Conseq} we get 
\ghostbox{$\proves d_0 : \spec{ an(n) }{ an(\elab(d_0,c,\fin)) }$}
and 
\ghostbox{$\proves d_1 : \spec{ an(n) }{ an(\elab(d_1,c,\fin)) }$}.
By definitions we have $\elab(b,c,\fin) = \elab(d_0,c,\fin) = \elab(d_1,c,\fin)$. 
So by rule \rn{Choice} we get 
\ghostbox{$\proves b : \spec{ an(n) }{ an(\elab(b,c,\fin) }$}.
\end{ditemize}

\subsection*{Proof of Theorem~\ref{thm:product}}


For basic semantics this is left to the reader.
We show that 
if $\Pi_{A,A'}$ is a strongly $\R$-adequate  product 
then,
in non-stuck semantics, 
$\Pi_{A,A'} \models \spec{\R}{\S}$ implies $A|A' \models \rspec{\R}{\S}$.

To show $A|A' \models \rspec{\R}{\S}$ in the non-stuck semantics of relational specs,
we must establish (i) all terminated traces of $A$ and $A'$ that initially satisfy $\R$
finally satisfy $\S$, and (ii) all finite and initial traces
of $A$ and $A'$ are not stuck.  Let $\tau, \tau'$ be finite initial traces of $A, A'$
such that $\tau_0, \tau'_0 \models \R$.
\begin{dlist}
\item[(i)]
  Assume $\tau$ and $\tau'$ are terminated, therefore $\ctrl(\last{\tau}) = \fin$ and
  $\ctrl(\last{\tau'}) = \fin'$.  Since $\Pi_{A,A'}$ is $\R$-adequate
  (implied by strong $\R$-adequacy), we can appeal to 
  the result for basic semantics to conclude
  $\last{\tau}, \last{\tau'} \models \S$. Hence, $\tau, \tau' \models \rspec{\R}{\S}$
  and we are done.
\item[(ii)]
  Assume $\tau$ is not terminated, i.e., $\ctrl(\last{\tau}) \neq \fin$.
  Since $\Pi_{A,A'}$ is strongly $\R$-adequate we can obtain a trace $T$ of
  $\Pi_{A,A'}$ such that $\tau \prefixeq \Left(T)$.  Now, there are two cases to consider.
  If $\tau \prefix \Left(T)$, then the witness for $\last{\tau} \trans -$ is the successor of
  $\last{\tau}$ in $\Left(T)$. Otherwise, $\tau = \Left(T)$.
  Since $\Pi_{A,A'}$ satisfies $\spec{\R}{\S}$ in the non-stuck semantics of unary
  specs, and is not terminated (because $\ctrl(\Left(\last{T}))=\ctrl(\last{\tau})$)
we know $T \biTrans -$.  This argument can be iterated, extending $T$, and 
after finitely many steps it must cover a left-side step from $\last{\tau}$ because otherwise it would contradict the one-side divergence condition of strong $\R$-adequacy of $\Pi_{A,A'}$.
The argument for $\tau'$ is similar, so we are done.
\end{dlist}

\medskip

Although for basic semantics the theorem is an equivalence, for non-stuck semantics the converse does not hold.
For a counterexample, consider the interleaved product (which is strongly $\R$-adequate) of $A,A'$ and
augment its control points to include a tag from $\{0,1\}$
(like the product in Sec.~\ref{sec:lockstep-seq}),
with tag value $1$ for the initial and final points.
The transition
relation is the same as that of the interleaved product, except it is not enabled for control points with
tag $0$.  Further, any step of the product may change the value of the tag nondeterministically.
Let $\Pi$ be such a product.
It is strongly $\R$-adequate.
But suppose $A|A' \models \rspec{\R}{\S}$, so there are no stuck non-final traces of $A$ or $A'$.
We do not have $\Pi \models \spec{\R}{\S}$.
If $T$ is a finite initial trace of $\Pi$ such that $T_0 \models \R$ and $\last{T}$ is not final,
we must show $T \biTrans -$.  Clearly, this cannot be true if the tag of $\ctrl(\last{T})$ is $0$; in this case, $T$ is stuck and $\last{T}$ has no successor.

\subsection*{Proof of Lemma~\ref{lem:product}}

By mutual implication.
For traces $\tau,\upsilon$ with $\upsilon_0=\last{\tau}$ we write
$\tau\diamond\upsilon$ for the coalesced catenation, i.e., $\tau$ followed by all but the first element of $\upsilon$. 

Assume $\models c;d' : \spec{\R^\eplus}{\S^\eplus}$.
To show $\models c|d : \rspec{\R}{\S}$,
consider terminated traces $\tau$ of $c$ and $\upsilon$ of $d$ (over $\Var$).  
Suppose $\tau_0,\upsilon_0\models\R$,
to show $\last{\tau}, \last{\upsilon} \models \S$.
By renaming we get a trace $\upsilon'$ of $d'$.  
Let $s=\store(\upsilon_0)\circ \fdot$.
By adding $s$ to the stores of $\tau$ we get a trace $\hat{\tau}$, of $c;d'$ 
that leaves the $\Dot{\Var}$ part unchanged and ends ready to execute $d'$.  
Adding $\store(\last{\tau})$ to the stores of $\upsilon'$ we get a trace $\hat{\upsilon}'$ of $d'$
that leaves the $\Var$ part unchanged.  
So $\hat{\tau}\diamond\hat{\upsilon}'$ is a terminated trace of $c;d'$ that ends
with store $\store(\last{\tau}) + \store(\last{\upsilon})$.
We have $\tau_0,\upsilon_0\models\R$ iff  $\tau_0\union s\models \R^\eplus$,
whence $(\hat{\tau}\diamond\hat{\upsilon}')_0\models\R^\eplus$.
So from $\models c;d' :  \spec{\R^\eplus}{\S^\eplus}$ 
we get $\last{(\tau\diamond\upsilon')}\models \S^\eplus$, 
that is, $\store(\last{\tau}) + \store(\last{\upsilon}) \models \S^\eplus$.
The latter is equivalent to 
$\store(\last{\tau}), \store(\last{\upsilon}) \models \S$
which completes the proof that $\tau,\upsilon\models\rspec{\R}{\S}$.

For the converse, assume $\models c|d :  \rspec{\R}{\S}$.
To show $\models c;d' : \spec{\R^\eplus}{\S^\eplus}$, 
consider any terminated trace of $c;d'$, which by semantics
can be written as coalesced catenation $\tau\diamond\upsilon$ of nonempty traces
of $c$ and $d'$, 
with the $\Dot{\Var}$-part of each $\tau_i$ equal to that of $\upsilon_0$,
and the $\Var$-part of each $\upsilon_i$ equal to that of $\last{\tau}$.
By discarding the untouched parts, and renaming, we get terminated $\Var$-traces 
$\hat{\tau}$ of $c$ and $\hat{\upsilon}$ of $d$.
Moreover if $\tau_0\models\R^\eplus$ then $\hat{\tau}_0,\hat{\upsilon}_0\models\R$ 
so using $\models c|d :  \rspec{\R}{\S}$ we get 
$\last{\hat{\tau}},\last{\hat{\upsilon}}\models\S$.
The latter is equivalent to $\last{\upsilon}\models\S^\eplus$
whence $(\tau\diamond\upsilon)\models\spec{\R^\eplus}{\S^\eplus}$.

\subsection*{Proof of Theorem~\ref{thm:lockstep-seq}}

As with the previous theorems, the associated judgments 
are easily determined by inspecting the proof of the theorem
so we refrain from spelling them out.

The proof relies on an analysis of the VCs.
These include VCs for the lockstep part of the automaton, like those in Fig.~\ref{tab:VClock},
as well as VCs for the one-sided part of the automaton like those in Figs.~\ref{tab:VCseq},
and we omit them. 
We use HL to obain proofs for $b$ and $\fdot(b')$, then lift those using \rn{SeqProd},
and complete the derivation of $\proves c|c' : \rspec{\R}{\S}$ using rules in Fig.~\ref{fig:lockstep}.
Here are more details.

By induction on $b$, using an argument similar to that proving (\ref{eq:seqA}), 
using left-side VCs, we have for all subcommands $d$ of $b$:
\begin{equation}\label{eq:mixA}
  \proves d : \spec{an(\lab(d),beg,lo)^\eplus}{an(\elab(d,c),beg,tag)^\eplus}
\end{equation}
where 
$tag=ro$ if $m=end$, else $tag=lo$.  (The different cases for $tag$ arise
from the form of the CFG, in particular transitions of type (iii) versus (iv).)
Then in particular the instantiation for $b$ itself is
$\proves b : \spec{an(beg,beg,lo)^\eplus}{an(end,beg,ro)^\eplus}$ 
with ending tag $ro$ because the paths through $b$ end with a transition of type (iv).
Next, by induction on $b'$, a similar argument using right-side VCs shows
for all subcommands $d'$ of $b'$:
\begin{equation}\label{eq:mixB}
\proves \fdot(d') : \spec{an(end,\lab(d'),ro)^\eplus}{an(end,m,tag)^\eplus} 
\end{equation}
where $m=\elab(d',c)$ and $tag=lck$ if $m=end$ else $tag=ro$.
Instantiating (\ref{eq:mixA}) and (\ref{eq:mixB}) we get 
\[ \begin{array}{l}
\proves b : \spec{an(beg,beg,lo)^\eplus}{an(end,beg,ro)^\eplus} \\
\proves \fdot(b') : \spec{an(end,beg,ro)^\eplus}{an(end,end,lck)^\eplus} 
\end{array}
\]
hence
$\proves b;\fdot(b') : \spec{an(beg,beg,lo)^\eplus}{an(end,end,lck)^\eplus}$
by \rn{Seq}. 
So rule \rn{SeqProd} yields
\begin{equation}\label{eq:mixin}
\proves b | b' : \rspec{an(beg,beg,lo)}{an(end,end,lck)}
\end{equation}
Now show, for all subcommands $d$ of $c$ and 
corresponding subcommands $d'$ of $c'$, 
with $d$ not a proper subcommand of $b$ (so $\lab(d')=\lab(d)$ and
$d'$ is not a proper subcommand of  $b'$):
\begin{equation}\label{eq:mixout}
\proves d \sep d' : \rspec{an(\lab(d),\lab(d),tag)}{an(m,m,tag_0)} 
\end{equation}
where $m=\elab(d,c,\fin)=\elab(d',c',\fin)$
and $tag=lo$ if $\lab(d)=beg$ (i.e., $d$ is $b$), and \(tag=lck\) otherwise;
and $tag_0=lo$ if $m=beg$, else $tag_0=lck$.
(Here the tag conditions are due to transitions (i) versus (ii);
noting that $\lab(d)=beg$ implies $m\neq beg$.)
The proof is by induction on structure of $c$ (which is structurally the same as $c'$).
The base cases $b|b'$, pairs of assignments outside $b,b'$, and pairs of skip.
For $b|b'$, (\ref{eq:mixout}) is just (\ref{eq:mixin}).
For assignments and skip, we use the VCs just as in the proof of Theorem~\ref{thm:lckcomplete}.
The induction step to prove (\ref{eq:mixout}) is same-structure command relations.
By analysis of the lockstep transitions of the product we get VCs like in Fig.~\ref{tab:VClock},
and using the assumption that the annotation's assertions imply test agreement on branches 
(outside $b$ and $b'$), we can show the rules in Fig.~\ref{fig:lockstep} suffice to prove what's needed.

Finally, instantiate $d,d'$ as $c,c'$ in (\ref{eq:mixout}),
noting that $\lab(d)=1\neq beg$ so $tag=lck$.
So Theorem~\ref{thm:lockstep-seq} is proved, since $an(1,1,lck) = \R$ and $an(\fin,\fin,lck)=\S$.

To handle the general case where there are multiple designated subprograms, disjoint from each other, one can define the product in terms of a set $BEG$ of begin labels and $END$ of end labels (disjoint from $BEG$), replacing condition $m\neq beg$ with $n\notin BEG$ and adjusting 
the other transition rules accordingly.

To avoid the case distinctions on tags in the
induction hypotheses (\ref{eq:mixA}) and (\ref{eq:mixB}),
one could relax the definition of product automata to allow a product
to take transitions with no effect on the underlying computation.  
In the present case we would use one 
from $(beg,beg,lck)$ to $(beg,beg,lo)$ and so forth.
For adequacy, such steps must be bounded. 
This is akin to the familiar requirement that ghost code must be terminating~\cite{FilliatreGP16}.

\subsection*{Exercise following Theorem~\ref{thm:lockstep-seq}}

Consider the situation assumed in Thm.~\ref{thm:lockstep-seq},
i.e., $\sameExcept(c,c',b,b',beg,end,\fin)$,
and suppose in addition that the distinguished subprograms $b$ and $b'$ are both choices, say
$\choice{b0}{b1}$ and $\choice{b0'}{b1'}$.
The product used in Thm.~\ref{thm:lockstep-seq} essentially executes 
$(\choice{b0}{b1});(\choice{b0'}{b1'})$ (ignoring the dot encoding, for clarity).
A proof constructed in accord with the theorem will use 
\rn{SeqProd} with a single relation $\Q$
at the semicolon.  Then using the HL rule for choice the judgments involved will be
$b0:\spec{\R}{\Q}$,
$b1:\spec{\R}{\Q}$,
$b0':\spec{\Q}{\S}$, and 
$b1':\spec{\Q}{\S}$.
The exercise is to modify the product to handle this situation
by using a four-way nondeterministic choice between sequentially executing
$b0;b0'$, $b0;b1'$, $b1;b0'$, or $b1;b1'$.
This corresponds to the relational four-premise rule for relating choices:
\[ \inferrule{
c\sep c' :\spec{\R}{\S}\\
c\sep d' :\spec{\R}{\S}\\
d\sep c' :\spec{\R}{\S}\\
d\sep d' :\spec{\R}{\S}
}{ (\choice{c}{d}) \sep (\choice{c'}{d'}) :\spec{\R}{\S} }
\]
(which is similar to the four-way rule relating if-else commands in some RHLs).
Then \rn{SeqProd} will be used for the judgments
$b0;b0':\spec{\R}{\S}$,
$b0;b1':\spec{\R}{\S}$,
$b1;b0':\spec{\R}{\S}$,
and 
$b1;b1':\spec{\R}{\S}$.
For these, we may choose four different intermediate assertions,
by contrast with a single $\Q$ that works for all four cases.

\subsection*{Theorem~\ref{thm:lockstep-While3}: proof and automaton}
Here is the automaton for the conditionally aligned loop product, under the conditions
given in Sec.~\ref{sec:dissonant}.
Note that $\labs(\sub(beg,c)) = \labs(\sub(beg,c'))$ under 
those conditions.

\begin{small}
\begin{list}{}{}
\item[(i)]
  $((n,n,lck),(s,s')) \biTrans ((m,m,lck),(t,t'))$ \\
  if $((n,n),(s,s')) \biTrans_{lckc} ((m,m),(t,t'))$ 
and moreover if $n=beg$ then $s,s'\not\models\Lrel$ and 
$s,s'\not\models\R$ 

\item[(ii)]
  $((beg,beg,lck),(s,s')) \biTrans ((m,beg,lo),(s,s'))$ \\
  if $(beg,s) \trans (m,s)$ and $s,s'\models \Lrel$
\item[(iii)]
  $((beg,beg,lck),(s,s')) \biTrans ((beg,m,ro),(s,s'))$ \\
  if $(beg,s') \trans' (m,s')$  and $s,s'\models\R$

\item[(iv)]
  $((n,beg,lo),(s,s')) \biTrans ((m,beg,lo),(t,s'))$ \\
  if $(n,s) \trans (m,t)$ and $m\neq beg$ and $n\in\labs(\sub(beg,c))$  

\item[(v)]
  $((beg,n,ro),(s,s')) \biTrans ((beg,m,ro),(s,t'))$ \\
  if $(n,s') \trans' (m,t')$ and $m\neq beg$ and $n\in\labs(\sub(beg,c))$  

\item[(vi)]
  $((n,beg,lo),(s,s')) \biTrans ((beg,beg,lck),(t,s'))$ \\
  if $(n,s) \trans (beg,t)$ and $n\in\labs(\sub(beg,c))$

\item[(vii)]
  $((beg,n,ro),(s,s')) \biTrans ((beg,beg,lck),(s,t'))$ \\
  if $(n,s') \trans' (beg,t')$ and $n\in\labs(\sub(beg,c))$ 
\end{list}
\end{small}

\begin{figure*}
\begin{small}
\begin{tabular}{llll}
CFG edge $(n,m,tag)\cfge\ldots$   & $\sub(n,c)$ and $\sub(m,c')$ & VC & rule  \\\hline
$(beg,beg,lck)\cfge(m,m,lo)$ & enter loop body jointly & $an(beg,beg,lck)\land\leftF{e}\land\rightF{e'}\imp an(m,m,lck)$ & (i) \\
$(beg,beg,lck)\cfge(m,beg,lo)$ & enter loop body left & $an(beg,beg,lck)\land\leftF{e}\land\Lrel\imp an(m,beg,lo)$ & (ii) \\
$(n,beg,lo)\cfge(m,beg,lo)$ & left if test $e$ true & $an(n,beg,lo)\land\leftF{e}\imp an(m,beg,lo)$ & (iv) \\
$(n,beg,lo)\cfge(beg,beg,lck)$ & end left iteration & $an(n,beg,lo)\imp an(beg,beg,lck)$ & (vi) \\
$(beg,beg,lck)\cfge(m,m,lck)$ & exit loop $m=\elab(...)$ & $an(beg,beg,lck)\land\neg\leftF{e}\land\neg\rightF{e'}\imp an(m,m,lck)$ & (i) \\
\end{tabular}
\end{small}
\caption{Selected VCs for the conditionally aligned loop product,
with reference to relevant transition rule.}\label{tab:VCwh3}
\end{figure*}

\begin{figure*}
\begin{footnotesize}
\[\begin{array}{lll}
\multicolumn{2}{l}{(n,n',tag) \cfge (m,m',t)\ldots  } & \mbox{the VC is}\ldots 
\\\hline
(1,1,lck) & (2,2,lck) & x=x'\land x>3 \:\imp\: \subst{(y=y'\land y>3)}{y,y'}{x,x'} \\
(2,2,lck)  & (3,3,lck) &  y=y' \land y>3 \:\imp\: \subst{(y=y' \land y>3 \land z>z'>0)}{z,z'}{24,16} \\
(3,3,lck) & (4,4,lck)  & y=y' \land y>3 \land z>z'>0 \:\imp\: \subst{\S}{w,w'}{0,0} \\
(4,4,lck) & (5,5,lck)  & \S \land y\neq 4 \land y'\neq 4 \land\neg\Lrel\land\neg\R
       \:\imp\: \S \land y>4 \land\neg\Lrel\land\neg\R \\ 
(5,5,lck) & (6,6,lck)  & \S\land y>4 \land w\mod 2 = 0 = w'\mod 3 
     \:\imp\: \S \land y>4 \land w\mod 2 = 0 = w'\mod 3  \\ 
(6,6,lck)  & (7,7,lck) & \S \land y>4 \land w\mod 2 = 0 = w'\mod 3 \:\imp\: 
                \subst{(\S \land y>4 \land w\mod 2 = 0 = w'\mod 3)}{z,z'}{z*y,z*2} \\
(7,7,lck) & (9,9,lck) &  \S \land y>4 \land w\mod 2 = 0 = w'\mod 3 \:\imp\: \subst{\S}{y,y'}{y-1,y-1} \\
(5,5,lck) & (8,8,lck)  & \S\land y>4 \land w\mod 2 \neq 0 \neq w'\mod 3 
              \:\imp\: \S \land y>4 \land w\mod 2 \neq 0 \neq w'\mod 3 \\ 
(8,8,lck) & (9,9,lck)  & \S \land y>4 \land w\mod 2 \neq 0 \neq w'\mod 3 \:\imp\: \S \\ 
(9,9,lck) & (4,4,lck) & \S \:\imp\: \S \\
(9,9,lck) & (0,0,lck) & \S \:\imp\: z>z' \\
(4,4,lck) & (5,4,lo) & \S \land y\neq 4 \land \Lrel \:\imp\: \S \land \Lrel  \\ 
(5,4,lo) & (6,4,lo) & \S\land\Lrel\land w\mod 2 = 0 \:\imp\: false \\ 
(5,4,lo)  & (8,4,lo) & \S \land \Lrel\land w\mod 2 \neq 0 \:\imp\: \S  \\ 
(8,4,lo) & (9,4,lo) & \S\:\imp\:\S  \\
(9,4,lo) & (4,4,lck) & \S\:\imp\: \S  
\end{array}\]
\end{footnotesize}
\caption{Selected VCs for the conditionally aligned loop product of $\aut(c4;\lskipc{\fin})$ and $\aut(c5;\lskipc{\fin})$
and the given annotation.
}\label{tab:VCc4c5}
\end{figure*}

Theorem~\ref{thm:lockstep-While3} is proved as follows,
with reference to the VCs in Fig.~\ref{tab:VCwh3}.
The VCs reflect that only same-value-test paths are taken, except in $lo$ or $ro$ iterations
under conditions $\Lrel$ and $\R$ respectively.
The omitted VCs include the symmetric right-only transitions and unreachable ones.
As an aside, Fig.~\ref{tab:VCc4c5} instantiates some VCs for the $c4,c5$ example of Sec.~\ref{sec:dissonant}.

Let $b$ be the body of the loop $\sub(beg,c)$
and $b'$ the body of the loop $\sub(beg,c')$ (it must be a loop, by $\sameCtl(c,c')$).

Labels have no significance on the proof rules and we will use $\skipc$ without a label 
in the following.  One could as well choose fresh labels so as to work entirely 
with labelled commands.

By induction on $b$ we show its relation with skip:
for all subcommands $d$ of $b$:
\begin{equation}\label{eq:wh3lbody}
\proves d \sep\skipc : \rspec{an(\lab(d),beg,lo)}{an(m,beg,tag)} 
\end{equation}
where $m=\elab(d,c,\fin)$
and $tag=lck$ if $m=beg$, else $tag=lo$.
For this purpose we use the left-side rules of Fig.~\ref{fig:one-side} together
with \rn{rConseq}, in an argument similar to proofs of the previous theorems.
For example, in case $d$ is an assignment $\lassg{n}{x}{e}$ we use
\rn{AssSkip} to get 
\[ \lassg{n}{x}{e}\sep\skipc : \rspec{\subst{an(m,beg,tag)}{x|}{e|}}{an(m,beg,tag)} \]
and then use \rn{rConseq} and the VC to strengthen $\subst{an(m,beg,tag)}{x|}{e|}$ to $an(n,beg,tag)$.

The relevant VCs arise from type (ii), (iv), and (vi) transitions
(see Fig.~\ref{tab:VCwh3}).
This is similar to the proof of Theorem~\ref{thm:seqcomplete} in the sense that 
the VCs are for one-side execution, but there we work with unary judgments and here 
we are using relational judgments (and relations are sets of store pairs, not sets of stores).

By induction on $b'$ we show its relation with skip:
for all subcommands $d'$ of $b'$:
\begin{equation}\label{eq:wh3rbody}
\proves \skipc\sep d' : \rspec{an(beg,\lab(d'),ro)}{an(beg,m,tag)} 
\end{equation}
where $m=\elab(d',c,\fin)$ 
and $tag=lck$ if $m=beg$, else $tag=ro$.
For this purpose we use the right-side rules which mirror those shown in Fig.~\ref{fig:one-side}.
The relevant VCs arise from type (iii), (v), and (vii) transitions.

By induction on $b$ we show its relation with $b'$:
for all subcommands $d$ of $b$ and corresponding subcommands $d'$ of $b'$:
\begin{equation}\label{eq:wh3jbody}
\proves d\sep d' : \rspec{an(\lab(d),\lab(d),lck)}{an(m,m,lck)} 
\end{equation}
where $m=\elab(d,c,\fin)$.
For this purpose we use the lockstep rules of Fig.~\ref{fig:lockstep}, similar to the proof of 
Theorem~\ref{thm:lckcomplete} and relying on assumption (a) of Theorem~\ref{thm:lockstep-While3}.

Instantiating (\ref{eq:wh3lbody}), (\ref{eq:wh3rbody}), and (\ref{eq:wh3jbody}),
noting that $\elab(b,c,\fin)=beg$, we get
\[ 
\begin{array}{l}
\proves b\sep\skipc : \rspec{an(\lab(b),beg,lo)}{an(beg,beg,lck)} \\
\proves \skipc\sep b' : \rspec{an(beg,\lab(b),ro)}{an(beg,beg,lck)} \\
\proves b\sep b' : \rspec{an(\lab(b),\lab(b),lck)}{an(beg,beg,lck)} 
\end{array}
\]
Now using \rn{rConseq} and the VCs we get 
\[ 
\begin{array}{l}
\proves b\sep\skipc : \rspec{an(beg,beg,lck)\land\leftF{e}\land\Lrel}{an(beg,beg,lck)} \\
\proves \skipc\sep b' : \rspec{an(beg,beg,lck)\land\rightF{e'}\land\R}{an(beg,beg,lck)} \\
\proves b\sep b' : \\ \rspec{an(beg,beg,lck)\land\leftF{e}\land\rightF{e'}\land\neg\Lrel\land\neg\R}{an(beg,beg,lck)} 
\end{array}
\]
With these, using assumption (b) of the theorem, we use rule \rn{caWhile} to get 
this judgment for the loop itself:
\begin{equation}\label{eq:wh3loop}
\begin{array}{l}
\sub(beg,c)\sep\sub(beg,c') :  \\
\quad \rspec{an(beg,beg,lck)}{an(beg,beg,lck)\land\neg\leftF{e}\land\neg\rightF{e'}}
\end{array}
\end{equation}
where $m=\elab(\sub(beg,c),c,\fin)$.

Now  by induction on $c$ we show for all subcommands $d$ of $c$
and corresponding subcommands $d'$ of $c'$,
such that $d$ is not a proper subcommand of 
the distinguished loop $\sub(beg,c)$ (nor $d'$ of $\sub(beg,c')$):
\begin{equation}\label{eq:wh3}
\proves d \sep d' : \rspec{an(\lab(d),\lab(d),lck)}{an(m,m,lck)} 
\end{equation}
where $m=\elab(d,c,\fin)$.
As base cases we have (\ref{eq:wh3loop}) as well as the atomic commands outside the distinguished loop $b$.
The induction steps are like those proving lockstep product
Theorem~\ref{thm:lckcomplete}.

Instantiating (\ref{eq:wh3}) with $c,c'$ we get 
\[ \proves c\sep c' : \rspec{an(1,1,lck)}{an(\fin,\fin,lck)} \]
As $an$ is an annotation for $\rspec{\P}{\Q}$, this is
$\proves c\sep c' : \rspec{\P}{\Q}$, q.e.d.

\medskip

Rule \rn{caWhile} essentially subsumes \rn{dWh} of Fig.~\ref{fig:lockstep},
in the presence of the rules of Fig.~\ref{fig:one-side}.
To see why, suppose we have the premise
$c\sep c' : \rspec{\Q\land \leftF{e}\land\rightF{e'}}{\Q}$ 
and side condition
$\Q \imp \eqbib{e}{e'}$ of \rn{dWh}.
Let $\Lrel=false=\R$.  
Then the side condition of \rn{caWhile},
$\Q \imp \eqbib{e}{e'} \lor (\Lrel\land\leftF{e}) \lor (\R\land\rightF{e'})$,
is equivalent to $\Q \imp \eqbib{e}{e'}$.
We can obtain the premises 
$c\sep\skipc : \rspec{\Q\land\Lrel\land \leftF{e}}{\Q}$ and
$\skipc\sep c' : \rspec{\Q\land\R\land \rightF{e'}}{\Q}$
of \rn{caWhile}, 
using \rn{rConseq}, from 
\[ c\sep\skipc : \rspec{false}{\Q} \quad\mbox{and}\quad
  \skipc\sep c' : \rspec{false}{\Q} \] 
and then use \rn{caWhile} to get the conclusion of \rn{dWh}.
Obviously  $c\sep c':\rspec{false}{\Q}$ is valid for any $c,c',\Q$.
Technically, to make \rn{dWh} a derived rule, we would need 
to add $c\sep c':\rspec{false}{\Q}$ as an axiom.

\subsection*{Soundness of \rn{caWhile}}

Soundness of a rule similar to \rn{caWhile} is proved by Beringer~\cite{Beringer11}
but since the rule is not widely known we sketch a proof here.

One way to prove soundness is by way of the conditionally aligned loop product.
A product gives rise to a simple proof structure, namely induction on traces.
For the sake of variety we sketch a different argument.

Suppose the premises hold:
\[\begin{array}{ll}
(i) & \models  c\sep c' : \rspec{\Q\land \leftF{e}\land\rightF{e'}\land\neg\Lrel\land\neg\R}{\Q} \\
(ii) & \models  c\sep\skipc : \rspec{\Q\land\Lrel\land \leftF{e}}{\Q} \\
(iii) & \models  \skipc\sep c' : \rspec{\Q\land\R\land \rightF{e'}}{\Q}
\end{array}
\]
Suppose the side condition 
$\Q \imp \eqbib{e}{e'} \lor (\Lrel\land\leftF{e}) \lor (\R\land\rightF{e'})$
is valid.  
To show validity of the conclusion,
$\models
  \whilec{e}{c} \Sep \whilec{e'}{c'} : \rspec{\Q}{\Q\land \neg\leftF{e}\land\neg\rightF{e'}}
$,
consider any terminated traces $\tau,\tau'$ of the left and right programs respectively, such that $\tau_0,\tau'_0\models\Q$.
We have $\last{\tau},\last{\tau'}\models \neg\leftF{e}\land\neg\rightF{e'}$ by 
program semantics. It remains to show 
$\last{\tau},\last{\tau'}\models\Q$.

We prove the following claim: 
for any terminated traces $\tau,\tau'$, if 
$\tau_0,\tau'_0\models\Q$ then $\last{\tau},\last{\tau'}\models\Q$.
The proof is by induction on $iter(\tau)+iter(\tau')$,
where $iter(\tau)$ is  the number of iterations on the left, and likewise for $iter(\tau')$ on the right.

The base case is $iter(\tau)+iter(\tau')=0$ in which case
$(\tau_0,\tau'_0) = (\last{\tau},\last{\tau'})$ and we are done.

Otherwise, suppose $iter(\tau)+iter(\tau') > 0$.
Then  $\tau(e)\neq 0$ or $\tau'(e')\neq 0$ since there is at least one iteration.  
So using the side condition
$\Q \imp \eqbib{e}{e'} \lor (\Lrel\land\leftF{e}) \lor (\R\land\rightF{e'})$
we have that $\tau_0,\tau_0'$ satisfies
\( (\leftF{e}\land\rightF{e'}) \lor (\Lrel\land\leftF{e}) \lor (\R\land\rightF{e'}) \).
This is equivalent to 
\[ (\leftF{e}\land\rightF{e'}\land\neg\Lrel\land\neg\R) \lor (\Lrel\land\leftF{e}) \lor (\R\land\rightF{e'}) \]
So it suffices to consider these three cases:
\begin{itemize}
\item $\tau_0,\tau_0'\models \leftF{e}\land\rightF{e'}\land\neg\Lrel\land\neg\R$

Both $\tau$ and $\tau'$ begin with an iteration;
let $\upsilon$ be the suffix of $\tau$ after its first iteration
and $\upsilon'$ the suffix of $\tau'$ after its first.
By premise $(i)$ we have $\upsilon_0,\upsilon'_0\models\Q$.
Now $iter(\upsilon)+iter(\upsilon') = iter(\tau)+iter(\tau')-2$ 
so we can apply the induction hypothesis to $\upsilon,\upsilon'$, which yields the result because $\last{\upsilon}=\last{\tau}$ and $\last{\upsilon'}=\last{\tau'}$.

\item  $\tau_0,\tau'_0\models \Lrel\land\leftF{e}$

So at least $\tau$ has an iteration.  Let $\upsilon$ be the suffix of $\tau$ after its first iteration.
By premise $(ii)$ we have $\upsilon_0,\tau'_0\models\Q$.
Now $iter(\upsilon)+iter(\tau') = iter(\tau) + iter(\tau') - 1 $
so we can apply the induction hypothesis to obtain the result.

\item $\tau_0,\tau'_0\models \R\land\rightF{e'}$

The argument goes like the second case but using $(iii)$.

\end{itemize}

\end{document}